\documentclass[epj,pacs]{svjour}
\usepackage{graphicx,amsmath,amssymb,bm}
\usepackage{color}
\usepackage[utf8]{inputenc}
\usepackage[english]{babel}
\usepackage[T1]{fontenc}
\usepackage{ragged2e}
\definecolor{purple}{rgb}{0.5,0,0.5}
\definecolor{blue}{rgb}{0.0,0,0.9}
\usepackage[colorlinks=true, pdfstartview=FitV, citecolor= purple, linkcolor = blue,urlcolor=blue]{hyperref}
\usepackage{amssymb}
\usepackage{lipsum}
\usepackage{graphics}
\usepackage{multirow}
\usepackage{epsfig}
\usepackage{rotating}
\usepackage{longtable} %\usepackage[english]{babel}
\usepackage{url}
\usepackage{pdflscape}
\usepackage{cancel}
\pdfoutput=1
\usepackage{csquotes}
\usepackage{placeins}
\usepackage{scalerel}
\usepackage{bbm}
\usepackage{adjustbox}
\usepackage{array}
\usepackage{rotating}
\newcommand{\Y}{\mathcal{Y}}
\usepackage{comment}
\usepackage{mathtools}
\usepackage{cuted}
\usepackage{caption}
\usepackage{widetext}

\begin{document}
\justifying
\title{Flavored axions and the flavor problem}

\author{Yithsbey Giraldo\inst{1} \thanks{yithsbey@gmail.com}\and R. Martinez\inst{2}\thanks{remartinezm@unal.edu.co} \and Eduardo Rojas\inst{1} \thanks{eduro4000@gmail.com} and Juan C. Salazar \inst{1}\thanks{jusala@gmail.com} 
}                     % Do not remove
\offprints{}          % Insert a name or remove this line
\institute{Departamento de F\'\i sica, Universidad de Nari\~no, A.A. 1175,  San Juan de Pasto, Colombia \and Departamento  de  Física,  Universidad  Nacional  de  Colombia, Ciudad Universitaria,  K.45 No.26-85, Bogota D.C., Colombia}
\date{Received: date / Revised version: date}

\abstract{A Peccei-Quinn~(PQ) symmetry is proposed, in order to generate in the Standard Model~(SM) quark sector a realistic mass matrix ansatz  with five texture-zeros. Limiting our analysis to Hermitian mass matrices we show that this requires a minimum of 4 Higgs doublets. This model allows assigning values close to 1 for several Yukawa couplings, giving insight into the origin of the mass scales in the SM. Since the PQ charges are non-universal the model features Flavor-Changing Neutral Currents~(FCNC) at the tree level. From the analytical expressions for the FCNC we report the allowed region in the parameter space obtained from the measurements of branching ratios of semileptonic meson decays.
}
\maketitle
%
%%%%%%%%%%%%%%%%%%%%%%%%%%%%%%%%%%%%%%%%%%%%%%%%%%%%%%%%%%%%%%%%%%%%%%%%%%%%%%%%%%%%%%
%%%%%%%%%%%%%%%%%%%%%%%%%%%%%%%%%%%%%%%%%%%%%%%%%%%%%%%%%%%%%%%%%%%%%%%%%%%%%%%%%%%%%%
\section{Introduction}
The discovery of the Higgs boson by the ATLAS~\cite{Aad:2012tfa} and CMS 
\cite{Chatrchyan:2012ufa}  collaborations with a mass of 125~GeV is very important because it opens up the possibility of new physics in the scalar sector. So that, from a theoretical viewpoint, 
an extended Higgs sector is well motivated~\cite{Gunion:1989we}, the best-known extensions are:  the two Higgs doublet model \cite{He:2008qm,Grzadkowski:2009iz,Logan:2010nw,Boucenna:2011hy,He:2011gc,Bai:2012nv,He:2013suk,Cai:2013zga,Wang:2014elb,Drozd:2014yla,Campbell:2015fra,vonBuddenbrock:2016rmr,Carena:2015moc} and models with additional singlet scalar 
fields~\cite{Branco:2011iw}.
On the other hand, the discovery of the Higgs boson gives experimental support to the
spontaneous symmetry breaking which is the mechanism that explains the origin 
of the masses for both, fermions and weak gauge bosons.

%of the mass of the fermions and the gauge bosons. 
The Standard Model (SM) symmetry breaking mechanism~\cite{Glashow:1961tr,Weinberg:1967tq,Salam:1968rm}  with Higgs-fermion couplings  proportional to the fermion masses
is consistent with  the experimental data;
%Using the Yukawa Lagrangian, the 
%Higgs mechanism allows the fermions to get masses.
however, there are various 
orders of magnitude between  the fermion masses which cannot be 
explained in the context of the SM.   %, 
These masses, the three mixing angles, and the complex CP-violating phase must be adjusted 
with experimental data.

The Two Higgs Doublet Model (THDM) was proposed in order to give masses to  up-type and
down-type quarks~\cite{Haber:1984rc} where vacuum expectation values (VEV) 
$v_1$ and $v_2$ are related to the electroweak VEV by the relation $v^2 = v_1^2 + v_2^2$.  This THDM allows new physics through an additional charged  scalar 
field which should be looked for in colliders as a test of multi-Higgs models.  On the other hand, the singlet scalar fields  are useful to break the $U(1)$ gauge 
symmetries in extended electroweak models or as candidates for dark 
matter~\cite{McDonald:1993ex,LopezHonorez:2006gr,Barbieri:2006dq,Honorez:2010re,Carena:2015moc}. 

Due to having three quarks up and three quarks down, the mass matrices are 
$3\times 3$, and under  a usual  assumption, these can be taken as Hermitian matrices having a total of 18 free parameters against the ten physical parameters
\cite{Ramond:1993kv}. 
This feature reduces the number of matrix parameters, easing the textures' analysis when comparing them with the experimental data.
One method to generate zeros and reduce the quark mass matrix parameters consists of performing a Weak Basis Transformation (WBT) on the quark fields~\cite{Fritzsch:1999ee,Branco:1988iq,Branco:1999nb}. By choosing these zeros by the WBT, the mass spectrum can be obtained according to the experimental results \cite{Branco:1988iq}. In particular, Fristzsch proposed a quark mass  matrix ansatz with six zeros\cite{Fritzsch:1977za,Fritzsch:1979zq,Du:1992iy,Fritzsch:1977vd}  which 
were put in by hand~\cite{He:1989eh}, but this texture predicted
for the ratio $|V_{ub}/V_{cb}|\approx 0.06$ a  too small magnitude~\cite{Beringer:1900zz}
which is  in strong tension with the present-day experimental result
$(|V_{ub} /V_{cb}|_{\text{exp}} \approx 0.09)$~\cite{Beringer:1900zz}. For this reason, some authors considered four zero-textures~\cite{Fritzsch:2002ga,Branco:1999nb,Xing:2003yj,Zhou:2003ji}. In reference~\cite{Giraldo:2011ya}, the matrices with  five 
texture-zeros could also explain the mass hierarchy and the parameters of the 
CKM matrix.

It is common to choose textures-zeros by hand without an underlying theory relying on first principles. Another direction that has been explored in the literature is to propose 
discrete symmetries and a sector with multiple scalar doublets to generate the 
textures of the quark mass matrices~\cite{Cheng:1987rs,DiazCruz:2004pj,Matsuda:2006xa,Carcamo:2006dp,Langacker:2000ju,Leroux:2001fx,Baek:2006bv,Ma:1995xk,Barger:2003zh,King:2005jy}. 
It is also possible to consider global symmetry groups  that prohibit 
certain Yukawa couplings and somehow generate the zeros of the mentioned 
textures~\cite{He:2006dk,Fukuyama:2010mz,Ahn:2013mva,Ferreira:2013oga,Felipe:2013ie,Ishimori:2012fg,Canales:2012dr,Kajiyama:2013sza,Mohapatra:2012tb,Varzielas:2012pa,Ding:2013hpa,Ma:2013xqa,Nishi:2013jqa,Ding:2013eca,Altarelli:2005yp,Ishimori:2010fs,Kadosh:2010rm,CentellesChulia:2020bnf}. 
\begin{comment}
Another way to get these textures is by means of a flavor dependent gauge symmetry, 
which can break the family universality of the Standard 
Model
\end{comment}
{Another way to obtain these textures is through a flavor-dependent gauge symmetry, which can break the family universality of the Standard Model}~\cite{Alvarado:2012xi,Carcamo:2006dp,Ibanez:1994ig,Binetruy:1994ru,Nir:1995bu,Jain:1994hd,Dudas:1995yu,Pisano:1991ee,Frampton:1992wt,Foot:1994ym,Hoang:1995vq,Hoang:1997su,Diaz:2002dx,Diaz:2003dk,Ochoa:2005ch}. 
\begin{comment}
This gauge 
symmetry produces textures which are linked to  additional flavor changing neutral 
currents and could, in principle, be measured in future colliders.
There are many proposed models with flavored gauge symmetries beyond the SM such 
as $SO(12)$, $SU(8)$, $331$, $U(1)$~\cite{Georgi:1979md,Barr:1979xt,Pati:1989ds,Barr:1989ta,Weinberg:1972ws,Georgi:1972mc,Mohapatra:1974wk,Balakrishna:1987qd,Balakrishna:1988ks,Barr:2001vj,Ferretti:2006df,Barr:2008pn,Frampton:2009ce,Babu:1995hr,Albright:1997xw,Sato:1997hv,Albright:1998vf,Irges:1998ax,Barr:2000ka,Haba:2000be,Babu:2001cv}, among others, which try to explain the flavor problem  
and the hierarchy of the SM masses.
\end{comment} 
{This gauge symmetry produces textures that are linked to additional flavor-changing neutral currents that, in principle, could be measured at future colliders. There are many proposed models with flavor gauge symmetries beyond the SM such as $SO(12)$, $SU(8)$, $331$, $U(1)$~\cite{Georgi:1979md,Barr:1979xt,Pati:1989ds,Barr:1989ta,Weinberg:1972ws,Georgi:1972mc,Mohapatra:1974wk,Balakrishna:1987qd,Balakrishna:1988ks,Barr:2001vj,Ferretti:2006df,Barr:2008pn,Frampton:2009ce,Babu:1995hr,Albright:1997xw,Sato:1997hv,Albright:1998vf,Irges:1998ax,Barr:2000ka,Haba:2000be,Babu:2001cv}, among others, that attempt to explain the flavor problem and the SM mass hierarchy.}
\begin{comment}
Alternative mechanisms to generate 
textures is through additional discrete global groups, i.e., $A_4$, $\Delta_{27}$, $Z_2$, $S_3$, 
etc
\end{comment} 
{Alternative mechanisms for generating textures are via additional discrete global groups, i.e., $A_4$, $\Delta_{27}$, $Z_2$, $S_3$, etc.}~\cite{He:2006dk,Fukuyama:2010mz,Ahn:2013mva,Ferreira:2013oga,Felipe:2013ie,Ishimori:2012fg,Canales:2012dr,Kajiyama:2013sza,Mohapatra:2012tb,Varzielas:2012pa,Ding:2013hpa,Ma:2013xqa,Nishi:2013jqa,Ding:2013eca,Altarelli:2005yp,Ishimori:2010fs,Kadosh:2010rm}. 
\begin{comment}
An interesting way used to 
explain the SM mass hierarchy is introducing exotic quarks with  ordinary charges 
that mix with the  ordinary  ones in the SM, producing small masses through seesaw mechanism
\end{comment} 
{An interesting way to explain the SM mass hierarchy is to introduce exotic quarks with ordinary charges that mix with the ordinary ones in the SM, producing small masses through the seesaw mechanism~\cite{Garnica:2019hvn}.} 

\begin{comment}
An important open problem  in particle physics is the strong CP violation 
associated with the $U(1)_A$ abelian symmetry~\cite{Weinberg:1975ui,tHooft:1976snw,Callan:1976je,tHooft:1986ooh} 
 which is restricted by  the constraints on the  neutron electric 
dipole moment~\cite{Baker:2006ts,Baluni:1978rf,Crewther:1979pi} 
 setting limits on the  $\theta$  parameter of 
the order of $10^{-10}$~\cite{Weinberg:1977ma,Wilczek:1977pj}. 
\end{comment} 
{An important open problem in particle physics is the strong CP violation associated with the abelian symmetry $U(1)_A$~\cite{Weinberg:1975ui,tHooft:1976snw,Callan:1976je,tHooft:1986ooh}, which is restricted by constraints on the electric dipole moment~\cite{Baker:2006ts,Baluni:1978rf,Crewther:1979pi} of the neutron that set limits on the $\theta$ parameter of the order of $10^{-10}$~\cite{Weinberg:1977ma,Wilczek:1977pj}.}
\begin{comment}
By introducing a 
chiral global symmetry or Peccei Quinn symmetry, this fine tuning can be 
explained.
\end{comment}
{
By introducing a global chiral symmetry or Peccei Quinn symmetry, this fine-tuning can be explained.}
\begin{comment}
But breaking this global symmetry implies the existence of a 
Goldstone boson which has been excluded by colliders and can not be coupled 
directly to the SM quarks. 
\end{comment}
{But breaking this global symmetry implies the existence of a Goldstone boson,}
\begin{comment}
this field is known as the axion and there are several 
models where the axion is invisible
\end{comment}
{This field is known as axion and there are several models in which the axion is invisible~\cite{Zhitnitsky:1980tq,Dine:1981rt,Mohapatra:1982tc,Shafi:1984ek,Langacker:1986rj,Shin:1987xc,Berezhiani:1989fp}.}
From cosmological 
considerations the axion decay constant $f_a$  must be of the order of $10^{7} 
- 10^{17}$~GeV. 
\begin{comment}
On the other hand, the axion acquires a mass different from 
zero due to the mixing with the $\pi^0$
and $\eta$ mesons, and
takes a mass given by \cite{Weinberg:1977ma,Stern:1970bh}
%
\begin{equation}
\label{eq1x}
m_a =\frac{\sqrt{m_um_d}}{m_u+m_d}\frac{m_\pi f_\pi}{f_a},
\end{equation}
%
where $m_\pi$ , $f_\pi$ denote the mass and decay constant of the pion and  $m_u$  and $m_d$  the up and down quark masses, respectively;
via this mixing, the axion  decays  into two photons. 
\end{comment}
{On the other hand, the axion acquires a non-zero mass due to mixing with the $\pi^0$ and $\eta$ mesons, and takes a mass given by~\cite{Weinberg:1977ma,Stern:1970bh}
\begin{equation}
\label{eq1x}
m_a =\frac{\sqrt{m_um_d}}{m_u+m_d}\frac{m_\pi f_\pi}{f_a},
\end{equation}
where $m_\pi$ , $f_\pi$ denote the mass and decay constant of the pion, and $m_u$ and $m_d$ the masses of the up and down quarks, respectively; by this mixing, the axion decays into two photons.}
\begin{comment}
The axion could also be a dark matter candidate  for values of the decay constant $f_a$ larger than  $10^{10}$ GeV, where the different production mechanisms of the axion field are~\cite{Preskill:1982cy,Abbott:1982af,Dine:1982ah}:  misalignment, global strings and domain
wall decays, etc, generating relic densities of the order of $0.12$. 
\end{comment}
{Axion could also be a dark matter candidate for values of the decay constant $f_a$ greater than $10^{10}$~GeV, where the different axion field production mechanisms are~\cite{Preskill:1982cy,Abbott:1982af,Dine:1982ah}: misalignment, global string and domain wall decays, etc, generating relic densities of the order of $0.12$.}
\begin{comment}
Experiments designed to study the decays $K^{\pm}\to\pi^{\pm} \nu\bar\nu$ are being reinterpreted to study decays with flavor changing through axions of the form $K^{\pm}\to\pi^{\pm} a$. 
\end{comment}
{Experiments designed to study $K^{\pm}\to\pi^{\pm} \nu\bar\nu$ decays are being reinterpreted to study flavor-changing decays through axions of the form $K^{\pm}\to\pi^{\pm} a$.}
\begin{comment}
Similarly, decays with flavor changing are studied in the bottom sector.  On the other hand, the effective coupling of the axion to  photons is excluded by low energy experiments and it must be less than $10^{-11}$.
\end{comment}
{Similarly, flavor-changing decays in the bottom  sector are studied.  On the other hand, the effective coupling of the axion to photons is excluded by low energy experiments and must be less than $10^{-11}$.}

The purpose of our work is to use the PQ symmetry to generate realistic mass textures that allow us to explain the quark masses and the CKM mixing matrix of the standard model and simultaneously the strong CP problem.
The idea of linking the PQ symmetry with the flavor problem was proposed in~\cite{Wilczek:1982rv}, and in later literature~\cite{Geng:1988nc,Berezhiani:1989fp,Hindmarsh:1997ac}.
Recently, there has been renewed interest in this direction~\cite{Bjorkeroth:2018dzu,Bjorkeroth:2018ipq,Reig:2018ocz,Linster:2018avp,Ahn:2018cau,Arias-Aragon:2017eww,Ema:2016ops,Calibbi:2016hwq,Alanne:2018fns,Bertolini:2014aia,Celis:2014jua,Ahn:2014gva,Cheung:2010hk,Albrecht:2010xh,Appelquist:2006xd}.
We impose a PQ symmetry on the SM, which can generate mass textures that reproduce the masses of the Standard Model quarks for Yukawa couplings close to unity. To obtain this result, a sector of multihiggs is needed in such a way that the hierarchy problem is reduced to defining the VEVs of the neutral components of the scalar doublets. 

This work is organized as follows:
In section~\ref{sec:georgi} we will summarize some results of the literature on five-zero textures, in section \ref{sec:III} we carry out an analysis of the PQ charges necessary to generate the textures of the quark mass matrices,  in this section, we also propose a natural way to normalize the PQ charges. In section~\ref{sec:IV} we will obtain the values of the vacuum expectation values VEV of the Higgs doublets to reproduce the masses of the quarks, in this section, we also determine the values of the Yukawa couplings and the minimum number of Higgs doublets necessary to generate the texture of the quark masses as shown in the Appendix~\ref{sec:demostracion}. In section~\ref{sec:effectiveL} we show the most general Lagrangian for the axion, and we calculate the masses of the scalar fields for typical values of  scalar potential couplings. In section~\ref{sec:constraints} we show the strongest constraints on the parameter space of the model. Finally in section~\ref{sec:conclusions} we present our the conclusions. 

%%%%%%%%%%%%%%%%%%%%%%%%%%%%%%%%%%%%%%%%%%%%%%
%%%%%%%%%%%%%%%%%%%%%%%%%%%%%%%%%%%%%%%%%%%%%%
\section{The five texture-zero mass matrices}
\label{sec:georgi}
One of the motivations to study the texture zeros in the Standard Model~(SM) 
and 
its extensions, is to simplify as much as possible the number of free parameters 
present in these models. The Yukawa Lagrangian, which is the responsible to give 
mass to the SM fermions after the spontaneous breaking of the electroweak 
symmetry $SU (2)_L\otimes U (1)_X \rightarrow U (1)_{EM}$, has 36 free parameters 
in the quark sector, enough to reproduce the experimental data in the literature, 
i.e.,  the 10 physical quantities in the quark sector~(6 quark masses,   3 
mixing angles and the CP violation phase of the  CKM matrix).
Without a Model to make predictions, discrete symmetries can be used to prohibit some components in the Yukawa matrix by generating the so-called texture zeros in the mass matrix. In many works instead of proposing a discrete symmetry, texture zeros are proposed as practical alternatives. This approach has as advantage that it is possible to choose the optimal mass matrix for analytical treatment of the problem, while simultaneously manage to adjust the mixing angles and quark masses.
In the literature there are many proposed five-zero textures for the SM quark 
mass 
matrices~\cite{Verma:2017ppl,Xing:2019vks,Fritzsch:1999ee,Desai:2000bu,Ludl:2015lta,Ponce:2013nsa} ~\footnote{The six-zero textures have already been 
ruled out because their predictions are outside the allowed experimental ranges. 
For a more detailed discussion see the references cited above.}. Several of 
these textures successfully reproduce the experimentally measured physical 
quantities. We chose the following five-zero texture because it gets a good fit 
for the quark masses and mixing parameters~\cite{Giraldo:2011ya,Giraldo:2015cpp,Giraldo:2018mqi}:
\begin{equation}
\label{5.1y}
\begin{split}
M^{U}&=
\begin{pmatrix}
 0&0&C_u\\
0&A_u &B_u\\
C_u^*&B_u^*&D_u
\end{pmatrix},
\\
M^{D}&=
\begin{pmatrix}
 0&C_d&0\\
C_d^*&0&B_d\\
0&B_d^*&A_d
\end{pmatrix},
\end{split}
\end{equation}
where $M^U$ and $M^D$  are the mass matrices for the up-type and 
down-type quarks, respectively.
Due to the mass matrices are Hermitian,  the off diagonal matrix elements are 
not
independent, hence, the number of texture zeros in both matrices sum 
five.  The  hermitian mass matrices has been widely employed by several authors~\cite{Xing:2015sva,Ramond:1993kv,Xing:2019vks,Ludl:2015lta,EmmanuelCosta:2009bx,Fritzsch:2002ga,Chiu:2000gw,Branco:1999nb}; however, the stability of this hypothesis under radiative corrections has been poorly studied. 
The stability of the texture-zeros under radiative corrections is guaranteed by the PQ symmetry; however, the stability of the Hermitian hypothesis deserves a separate study as it is pointed out in reference~\cite{Xing:2015sva}. In such a reference, the authors concluded that the studied texture zeros of $M_u$ and $M_d$  are essentially stable against the evolution of energy scales in an analytical way by using the one-loop renormalization-group equations.
By using a WBT~\cite{Giraldo:2011ya,Branco:1988iq,Branco:1999nb} 
it is possible to remove the phases in $M^{D}$ to be absorbed by $M^{U}$,  i.e., 
 the phases in $B_d$ and $C_d$ are absorbed in $B_u$ and $C_u$, so that the mass 
matrices~\eqref{5.1y}  can be rewritten as:
{\footnotesize
\begin{equation}
\label{eq2}
\begin{split}
M^{U}&
=
\begin{pmatrix}
 0&0&|C_u|e^{i\phi_{C_u}}\\
0&A_u &|B_u|e^{i\phi_{B_u}}\\
|C_u|e^{-i\phi_{C_u}}&|B_u|e^{-i\phi_{B_u}}&D_u
\end{pmatrix},
\\
M^{D}&=
\begin{pmatrix}
 0&|C_d|&0\\
|C_d|&0&|B_d|\\
0&|B_d|&A_d
\end{pmatrix},
\end{split}
\end{equation}}
where $\phi_ {B_u} $ and $\phi_ {C_u} $ are the respective phases of the complex entries $ B_u$ and $C_u $. 
Since the trace and the determinant of a matrix are invariant under the diagonalization process, we can compare  these invariants for the mass matrices~\eqref{eq2} with the corresponding expressions in the mass basis where these matrices are diagonal, in such a way that we can write down the free parameters of $M^{U}$ and $M^{D}$  in terms of the quark masses.
\begin{subequations}
\label{e3.4}
\begin{align}
\label{3.18}
 D_u&=m_u-m_c+m_t-A_u,\\
\label{34a}
|B_u|&=\sqrt{\frac{(A_u-m_u)(A_u+m_c)(m_t-A_u)}{A_u}},\\
\label{35a}
|C_u|&=\sqrt{\frac{m_u\,m_c\,m_t}{A_u}},
\\
 A_d&=m_d-m_s+m_b,\\
\label{34b}
|B_d|&=\sqrt{\frac{(m_b-m_s)(m_d+m_b)(m_s-m_d)}{m_d-m_s+m_b}},\\
\label{35b}
|C_d|&=\sqrt{\frac{m_d\,m_s\,m_b}{m_d-m_s+m_b}}.
\end{align}
\end{subequations}
For reasons of convenience we have imposed that the eigenvalues of the mass matrices 
for the second generation take the negative values $- m_c $ and $- m_s $.
$ A_u $ is left as a free parameter and its value, determined  by  the hierarchy of the quark masses,  must be in the following interval:
\begin{equation}
\label{eq5}
 m_u\le A_u\le m_t.
\end{equation}
The exact analytical diagonalization mass matrices in Eq.~\eqref{eq2} are shown in Appendix~\ref{sec:mat-diag}.

%%%%%%%%%%%%%%%%%%%%%%%%%%%%%%%%%%%%%%%%%%%%%%%%%%%%%%%%%%%%%%%
%%%%%%%%%%%%%%%%%%%%%%%%%%%%%%%%%%%%%%%%%%%%%%%%%%%%%%%%%%%%%%%
\section{Textures, PQ symmetry and the minimal particle content}
\label{sec:III}
The five-texture zeros present in the mass matrices~\eqref {5.1y} can be generated through a 
 PQ symmetry  $U(1)_{PQ}$ on the Yukawa interaction terms between the SM fermions and the scalar doublets $\Phi^{\alpha}$ in the model~\cite{Bjorkeroth:2018ipq,Garnica:2019hvn,Ringwald:2015dsf}.
%%%%%%%%%%%%%
We also included  a heavy neutral quark $Q$, and  two scalar singlets $S_1$ and $S_2$;  the heavy quark is required to avoid the FCNC constraints while keeping the QCD anomaly at a finite value,  as it will be explained below.  The scalar singlet $S_1$ is necessary to break the PQ symmetry down at a given high energy scale $\Lambda_{PQ}$ (In principle, $S_2$ also breaks the PQ symmetry; however;  the purpose of $S_2$ is to give mass to the heavy quark, $S_1$ cannot give mass to the heavy quark due to its PQ charge). 
%%%%%%%%%%%%%%
The  Leading Order~(LO) Lagrangian for these fields is given by~\cite{Brivio:2017ije}: 
\begin{align}
\mathcal{L}_{\text{LO}}& \supset 
(D_\mu\Phi^{\alpha})^\dagger D^\mu\Phi^{\alpha}
+\sum_{\psi}i\bar{\psi}\gamma^{\mu}D_\mu \psi%\notag\\
+\sum_{i=1}^{2} (D_\mu S_i)^\dagger D^\mu S_i\notag\\
&%\frac{1}{2}\partial_{\mu}a\partial^{\mu}a-\frac{1}{2}m_a^2a^2
- \Bigg(
\bar{q}_{Li}y^{D\alpha}_{ij}      \Phi^{\alpha}d_{Rj}  +
\bar{q}_{Li}y^{U\alpha}_{ij}\tilde\Phi^{\alpha}u_{Rj} \notag\\
&+
\bar{\ell}_{Li}y^{E\alpha}_{ij}  \Phi^{\alpha}e_{Rj}+
\bar{\ell}_{Li}y^{N\alpha}_{ij}\tilde\Phi^{\alpha}\nu_{Rj}  +\text{h.c} \Bigg)\notag\\
+&(\lambda_Q\bar{Q}_R Q_LS_2+\text{h.c})-V(\Phi,S_1,S_2),%\notag\\
%\mathcal{L}& \supset 
%- \left(
%\bar{q}_{Li}y^{D\alpha}_{ij}      \Phi^{\alpha}d_{Rj}  +
%\bar{q}_{Li}y^{U\alpha}_{ij}\tilde\Phi^{\alpha}u_{Rj}  +
%\bar{l}_{Li}y^{E\alpha}_{ij}      \Phi^{\alpha}e_{Rj}
%\text{h.c} \right),
\label{eq6}
\end{align}
%%%%%%%%%%%%%%%%%%%%%%%%%%%%%%%%%%%%%%%%
 As it is shown in the Appendix~\ref{sec:demostracion}, the minimum number of Higgs doublets necessary to generate the texture of the quark masses is four, hence $\alpha=1,2,3,4$.
In this expression $i,j$ are family indices~(there is an implicit sum over repeated indices),
the superindex  $U$ refers to  up-type  quarks~(the same is true for the super 
indices $D$, $E$, $N$
which refer to down-type quark, electron-like and neutrino-like fermions, 
respectively) and
 $D_\mu =\partial_\mu+i\Gamma_\mu$ is the covariant derivative in the SM.
%%%%%%%%%%%%%%%%%%%%%%%%%%%%%%%%%%%%%%%%
$V(\Phi,S_1,S_2)$ is the scalar potential which  is shown in the Appendix~\ref{sec:scalars}.
In Eq.~\eqref{eq6}  $\psi$  stands for the standard model fermion fields plus the heavy quark $Q$.
As it is shown in Table~\ref{tab:pcontent2} the PQ charges of the heavy quark can be chosen in such a way that only the  interaction with the scalar singlet $S_2$ is allowed. 
%
%%%%%%%%%%%%%%%%%%%%%%%%%%%%%%%%%%%%%%%%%%%%%%%%%%
In our approach we assign  charges $Q_{\text {PQ}}$ to the quark sector particles for the left-handed doublets~($q_L$): $x_{q_i}$, up-type right-handed singlets~($u_R $): $x_{u_i}$ and down-type right-handed singlets~($ d_R $): $x_{d_i}$ for each family~($ i = 1,2,3 $), for the scalar doublets, $x_{\phi_{\alpha}}$~ ($\alpha=1,2,3,4$) and for the scalar singlets $x_{_{S_{1,2}}}$. 
For the time being  we only consider the quark sector but a similar analysis can be done in the lepton sector~\cite{Benavides:2020pjx}.
To forbid a given entry in the quark mass matrix, the corresponding sum of the PQ charges for the Yukawa interaction terms must be different from zero, i.e.,    $(-x_{q_i}+x_{u_j}-x_{\phi_\alpha})\neq 0$,  so that we can obtain texture-zeros by imposing the following conditions:

 \begingroup
\renewcommand*{\arraystretch}{1.5} 
\begin{align}\label{eq:conditions}
& M^{U}=
\begin{pmatrix}
0&0&x\\
0&x&x\\
x&x&x
\end{pmatrix}
\longrightarrow 
\begin{pmatrix}
 S_{11}^{U}\neq 0 & S_{12}^{U}\neq 0 & S_{13}^{U}= 0\\
 S_{21}^{U}\neq 0 & S_{22}^{U}   = 0 & S_{23}^{U}= 0 \\
 S_{31}^{U}=    0 & S_{32}^{U}   = 0 & S_{33}^{U}= 0 
\end{pmatrix},\notag
\\
&
M^{D}=
\begin{pmatrix}
0&x&0\\
x&0&x\\
0&x&x
\end{pmatrix}
\longrightarrow 
\begin{pmatrix}
 S_{11}^{D}\neq 0 & S_{12}^{D}    =0 & S_{13}^{D}\neq 0\\
 S_{21}^{D}    =0 & S_{22}^{D}\neq 0 & S_{23}^{D}    =0 \\
 S_{31}^{D}\neq 0 & S_{32}^{D}    =0 & S_{33}^{D}    =0 
\end{pmatrix},
\end{align}
\endgroup
where
\begin{align}
S_{ij}^{U}=& (-x_{q_i}+x_{u_j}-x_{\phi_\alpha}),\notag\\
S_{ij}^{D}=& (-x_{q_i}+x_{d_j}+x_{\phi_\alpha}) .    
\end{align}
In the matrix elements of the~Eq.~\eqref{eq:conditions} every equality must be satisfied only by one of the Higgs doublets, so in principle, we have  11 equations. The inequalities must be satisfied by all the Higgs charges  $x_{\phi_\alpha}$  therefore we have $7\times 4$ inequalities.
%In Appendix~\ref{sec:demostracion}, it is shown that the analyzed texture~\eqref{5.1y} needs at least four Higgs doublets, i.e., $n=4$,  to reproduce the five-texture zeros with the PQ symmetry $U (1)_{\text{PQ}}$ 
%
We will use the parametrization shown in the Tables~\ref{tab:pcontent1} and \ref{tab:pcontent2}.
The scalar singlets, $S_{1,2}$,  acquire a vacuum expectation value  at very high energies, where the PQ symmetry  is broken. 
Higgs doublets $\Phi^{\alpha}$ adquire VEVs around the electroweak scale.
Due to the particular choice of the  PQ charge for the scalar singlet $S_1$ (with a VEV of order $10^6$~GeV), trilinear terms,  coupling the scalar singlet  
$S_1$ to the scalar doublets $\Phi_\alpha$,  are allowed in the scalar potential $V(\Phi,S_1,S_1)$ (see Appendix~\ref{sec:scalars}), which are useful to have a spectrum of heavy scalar doublets above the TeVs.
The scalar masses are above the searches for heavy-neutral Higgs bosons for the typical benchmark models reported by ATLAS and CMS collaborations~\cite{Workman:2022ynf}.

The scalar potential  $V(\phi_\alpha,S_1,S_2)$ is invariant under the symmetry  $S_2\longrightarrow S_2^{\dagger}$~(which is equivalent to a $Z_2$ symmetry), but this symmetry is broken by the interaction term $\lambda_Q \bar{Q}_RQ_L S_2+\text{h.c.}$. 
In fact, from this interaction,  it is also possible to generate, at one loop, a mass term for the CP-odd field 
%$ S_i=\frac{v_{S_i}+\xi_{S_i}+i\zeta_{S_i}}{\sqrt{2}};\hspace{1cm} i=1,2.$
$\frac{1}{2} \left(m_{\zeta_{S_2}}\right)^2_{\text{SB}}\zeta^2_{S_2}$  in the effective Weinberg-Coleman 
potential (where $\zeta_{S_2}$ is the imaginary part of $S_2$)  
From this interaction, there is also a self-energy correction for CP-even fields, but it comes in with an opposite sign, so these corrections softly break the $Z_2$ symmetry.  As a consequence of this, $\zeta_{S_2}$ acquires a mass in the broken phase~\cite{Barger:2008jx,Barger:2010yn,Gross:2017dan,Chiang:2017nmu,Kannike:2019mzk,Abe:2020iph}.
From Eq.~\eqref{eq:scalar-potential} of Appendix~\ref{sec:scalars} it is possible to obtain the decay of Im$S_2=\zeta_{S_2} $ in two axions which depends on the parameter $\lambda_{S_1S_2}$,  $\zeta_{S_2}$ can also decay in
in two SM Higgs bosons, from the term $\sum_i \lambda_{iS_2} \Phi_i^{\dagger} \Phi_i S_2^{*}S_2$, therefore, their interactions are not well constrained by colliders, 
the impact on the parameter space of our model from the cosmological signatures of this scalar is beyond the purpose of the present work and deserves a dedicated studio.
%\vspace{-0.5cm}
\begin{strip}
\centering
\begin{tabular}{ccccccccc}
\hline  
\hline 
Particles & Spin &$SU(3)_C$ &$SU(2)_L$ &   $U(1)_Y$&    $U(1)_{\text{PQ}}$  &$Q_{\text{PQ}}(i=1)$& $Q_{\text{PQ}}(i=2)$& $Q_{\text{PQ}}(i=3)$\\
\hline
$q_{Li}$  & 1/2  & 3        &  2  & 1/6& $x_{q_i}$ &  $-2 s_1 + 2 s_2 + \alpha$  & $-s_1 + s_2 + \alpha$   & $\alpha$        \\  
$u_{Ri}$  & 1/2  & 3        &  1  & 2/3 & $x_{u_i}$     &  $s_1 + \alpha$ &  $s_2 + \alpha$  &  $-s_1 + 2 s_2 + \alpha$   \\
$d_{Ri}$  & 1/2  & 3        &  1  & -1/3  & $x_{d_i}$&  $2 s_1 - 3 s_2 + \alpha$ &  $s_1 - 2 s_2 + \alpha$  & $-s_2 + \alpha$  \\
\hline
\hline 
\end{tabular}
\captionof{table}{The columns 6-8 are the PQ~($Q_{PQ}$) charges for the SM quarks in each family. The subindex  $i=1,2,3$ stands for the family number in the interaction basis.  The parameters $s_1, s_2$ and $\alpha$ are reals, with 
$s_1\ne s_2$. }
\label{tab:pcontent1}
%\end{table}
\end{strip}
\begin{widetext}
%\begin{strip}
\begin{table}[h]
\begin{center}
%\bgroup                    % nuevo    
%\def\arraystretch{1.3}% nuevo
\begin{adjustbox}{max width=0.7\textwidth}
\begin{tabular}{ccccccccc}
\hline  
\hline  
Particles & Spin &$SU(3)_C$ &$SU(2)_L$ &$U(1)_Y$&   $U(1)_{\text{PQ}}$ &$Q_{\text{PQ}}$ \\
\hline
$\Phi_{1}$  & 0  & 1        &  2       &   1/2  &  $x_{\phi_1}$ & $s_1\in\text{reals}$  \\  
$\Phi_{2}$  & 0  & 1        &  2       &   1/2  &  $x_{\phi_2}$  & $s_2\in\text{reals}$ \\  
$\Phi_{3}$  & 0  & 1        &  2       &   1/2&  $x_{\phi_3}$   & $-s_1 + 2 s_2$    \\
$\Phi_{4}$  & 0  & 1        &  2       &   1/2 &  $x_{\phi_4}$  & $-3 s_1 + 4 s_2$  \\  
$S_1$       & 0  & 1        &  1       &    0  &  $x_{_{S_1}}$     & $x_{_{S_1}}=s_1-s_2\neq0$                 \\ 
$S_2$       & 0  & 1        &  1       &    0  &  $x_{_{S_2}}$      & $x_{_{S_2}}=x_{Q_R}-x_{Q_L}\neq0$                \\ 
$Q_L$       & 1/2& 3        &  0       &    0  &  $x_{Q_L}$   & $x_{Q_L}\in\text{reals}$ \\ 
$Q_R$       & 1/2& 3        &  0       &    0   &  $x_{Q_R} $  & $x_{Q_R}\in\text{reals}$    \\  
\hline
\hline  
\end{tabular}
\end{adjustbox}
%\egroup 
\caption{Beyond SM scalar and fermion fields and their respective PQ charges.
 The parameters $s_1, s_2$ are reals, with 
$s_1\ne s_2$.}
\label{tab:pcontent2}
%} 
\end{center}
\end{table}
%\end{strip}
\end{widetext}
%\begin{table}[h]
%
%$s_1\ne s_2$, where: $s_1= \frac{N}{9}\hat{s}_1$  and 
%$s_2=\frac{N}{9}\left(\epsilon+\hat{s}_1 \right)=\frac{N}{9}\hat s_2$  and  
% In order to account for  the parameter space of the PQ charges \textcolor{red}{No entiendo esta frase, para dar cuenta el espacio de parametros...}, 
% it is necessary to impose some normalization on the charges. 
% In our approach  the parameter space of the PQ charges  is 
%a tridimensional space, hence, 
%following a similar procedure as in $E_6$~\cite{London:1986dk,Erler:2011ud,Rojas:2015tqa,Benavides:2018fzm}, it will be advantageous to write the PQ charges as a linear combination
%of  conveniently chosen PQ charges associated with the $U(1)_{PQ}$ symmetries.
  As it is usual in the PQ formalism, we are interested in those charges for which the QCD anomaly $N$ is different from zero, where
\begin{align}
\label{eq:parametrization2}
N= 2\sum_{i}^3x_{q_i}-\sum_{i}^3 x_{u_i}-\sum_{i}^3 x_{d_i}+A_{Q},
\end{align}  
for this reason, in the literature $N$ is used as the normalization of the Peccei-Quinn~(PQ) charges. 
In order to generate the proper normalization to the charges in the Tables~\ref{tab:pcontent1} and ~\ref{tab:pcontent2}, for an SM fermion $\psi$ the most general 
PQ charges that reproduce the texture in Ref.~\cite{Giraldo:2011ya} are given by 
the  parametrization
\begin{eqnarray}
Q_{PQ}(\hat{s}_1,\epsilon,N,\alpha)(\psi)&=&
\frac{N}{9}\left(\hat{s}_1 Q^{s_1}_{PQ}(\psi)\right.
+ \left.(\epsilon+\hat{s}_1) Q^{s_2}_{PQ}(\psi)\right)\nonumber\\
&&+\alpha Q^{V}_{PQ}(\psi).
\label{eq:parametrization}
\end{eqnarray}
In this  expression, $ Q^{s_1,s_2,V}_{PQ}(\psi)$ are PQ charges, 
whose explicit expressions are given in the Table~\ref{tab:models}, where
$N= x_{Q_L}-x_{Q_R}+s_2-s_1$, $\epsilon=(1-A_Q/N)$, $\hat{s}_{1,2}=\frac{9}{N}s_{1,2}$ (are arbitrary real numbers such that $\hat{s}_1\neq \hat{s}_2$) and  $A_{Q}=x_{Q_{L}}-x_{Q_{R}}$  is the contribution to the anomaly  
of the heavy   quark $Q$, which is a singlet under the electroweak gauge group,
with  left\,(right)-handed Peccei-Quinn charges denoted by $x_{Q_{L}}(x_{Q_{R}})$.
This parametrization was obtained from  Tables~\ref{tab:pcontent1}  and \ref{tab:pcontent2}, by normalizing the PQ charges  in such a way that  the anomaly of $SU(3)_C$ is $N$ for any real value of the parameters  $\hat{s}_1$, $\hat{s}_2$, $\alpha$, $x_{Q_L}$ and  $x_{Q_R}$.
Because  $x_{Q_L}$ and $x_{Q_R}$ always appear in the combination $x_{Q_L}- x_{Q_R}=N(1-\epsilon)$
and $\hat{s}_2=\hat{s}_1+\epsilon$,
it is more convenient to use the set of parameters $\hat{s}_1$, $\epsilon$, $N$ and $\alpha$. 
For the FCNC processes considered in the present work,  the phenomenological couplings are proportional to differences between the PQ charges of down-type quarks, so that only $\epsilon$ and $N$ are relevant for these observables. 
To solve the strong CP problem $N\ne 0$  and to generate the texture-zeros in the mass matrices it is necessary to keep $\epsilon\ne 0$. 
It is important to note that due to the exotic heavy quark $Q$, It is possible to choose small PQ charges for SM fermions with a small contribution to the QCD anomaly, while the QCD anomaly remains finite (this condition is necessary to solve the strong CP problem), small couplings are also  important  to avoid collider constraints.   
\begin{table}[h]
\begin{center}
{\begin{tabular}{cccccccccc}
\hline 
\hline
Particles &\multicolumn{3}{c}{$ Q^V_{PQ}$}  &\multicolumn{3}{c}{$Q^{s_1}_{PQ}$} &\multicolumn{3}{c}{$Q^{s_2}_{PQ}$} \\
\hline
$q_{Li}$  & 1 & 1 & 1     & $-2$ & $-1$ &  0     & 2   &  1   & 0        \\  
$u_{Ri}$  & 1 & 1 & 1     &  1   &  0   & $-1$   & 0   &  1   & 2        \\
$d_{Ri}$  & 1 & 1 & 1     &  2   &  1   &  0     &$-3$ & $-2$ & $-1$     \\
\hline
Family       &1  & 2 & 3     &  1 &  2  & 3     &  1 &  2  & 3    \\
\hline
\hline
\end{tabular}
%\egroup
\caption{ The three columns in each slot are the Peccei-Quinn  charges $x_{\psi_i}$ in each family. 
PQ charges are shown for the SM left-handed quarks $x_{q_i}$, the right-handed up-type $x_{u_i}$ and down-type $x_{d_i}$ quarks for the three families  $i=1,2,3$.}
%} 
\label{tab:models}
}
\end{center}
\end{table}

The QCD anomaly is also  given by $N=A_Q/(1-\epsilon)$,
this parametrization is quite convenient since by fixing $N$ and $f_a$ for FCNC observables (for which $\alpha$ does not matter) in Eq.~\eqref{eq:parametrization} we can vary $\hat{s}_1$ and $\epsilon$ for a fixed $\Lambda_{PQ}=f_a N $, in such a way that the parameter space is naturally reduced to two dimensions.

If we want to solve the domain-wall problem  is necessary to calculate the QCD anomaly in a normalization such that the minimum magnitude of the non-vanishing PQ charges of the scalar fields and the quark condensates is 1~\cite{Kim:1986ax}.
The anomaly is given by $N=x_{Q_L}-x_{Q_R}+s_2-s_1$,  by choosing $s_2=-1$, $s_1=0$ and $\alpha=0$. In this case the charge of the singlet scalar $S_1$ is  $s_1-s_2=1$.  In this normalization, we can identify $N$ with  $N_{DW}$~\cite{Kim:1986ax} in such a way that  $N=N_{DW}=1$, which is equivalent to  $\epsilon=(x_{Q_L}-x_{Q_R})/N=2$.  There are other ways of choosing the parameters which also solve the problem.
The DW problem can be disposed of by introducing an explicit breaking of the PQ symmetry so that the degeneracy between the different vacua is removed and there is a unique minimum of the potential~\cite{DiLuzio:2020wdo}.

%%%%%%%%%%%%%%%%%%%%
%%%%%%%%%%%%%%%%%%%%%%
\section{The effective  lagrangian}
\label{sec:effectiveL}
The most important phenomenological consequence of  non-universal PQ charges 
is  the presence of FCNC.  To determine the restrictions coming from the FCNC we start by writing  the most general effective Lagrangian as~\cite{Georgi:1986df,Gavela:2019wzg}:
\begin{eqnarray}\label{eq:lagrangian}
\mathcal{L}_{\text{NLO}}&=
+c_{a\Phi^\alpha}O_{a\Phi^\alpha}+c_1\frac{\alpha_1}{8\pi}O_{B}\notag\\
&+c_2\frac{\alpha_2}{8\pi}O_{W}
+c_3\frac{\alpha_3}{8\pi}O_{G},
\end{eqnarray}
 $c_{a\Phi^\alpha}$ and $c_{1,2,3}$ are Wilson coefficients; $\alpha _{1,2,3}= 
\frac{g_{1,2,3}^2}{4\pi}$  where the $g_{1,2,3}$ are the coupling strengths of 
the electroweak interaction in the interaction basis; $q_{Li}$, $d_{Ri}$ and  
$u_{Ri}$, are the left-handed quark doublet, right-handed down-type   and right-handed up-type quark fields, 
respectively; $\ell_{Li}$, $e_{Ri}$ and $\nu_{Ri}$
 are the left-handed lepton doublet, right-handed charged lepton,  and right-handed neutrino fields, respectively.
 $\psi$ stands for the SM fermion fields  and the effective operators are given by
 \begin{align}
O_{a\Phi^\alpha}=& i\frac{\partial^\mu a}{\Lambda}   
\left((D_{\mu}\Phi^\alpha)^{\dagger} \Phi^\alpha-\Phi^{\alpha\dagger}(D_{\mu}\Phi^\alpha)\right),\notag\\
%         =&\frac{2i\partial^{\mu}a}{f_a}(D_\mu \Phi)^\dagger      \Phi + \frac{i\square a}{f_a}\Phi^{\dagger}\Phi
%         =-\frac{2i\partial^{\mu}a}{f_a}        \Phi^\dagger D_\mu\Phi - \frac{i\square a}{f_a}\Phi^{\dagger}\Phi\\
   O_{B}= & -\frac{ a}{\Lambda}B_{\mu\nu}\tilde B^{\mu\nu},\notag\\
   O_{W}=& -\frac{ a}{\Lambda}W_{\mu\nu}^a\tilde W^{a\mu\nu},\notag\\
   O_{G}=&  -\frac{ a}{\Lambda}G_{\mu\nu}^a\tilde G^{a\mu\nu},
\end{align}
where $B$,  $W^a$  and $G^a$ correspond to the gauge fields associated with the SM gauge groups  $U(1)_Y$,  $SU(2)_L$  
and  $SU(3)_C$, respectively. 
Redefining the fields~\cite{Georgi:1986df}
\begin{align}
\label{eq:redefinitions}
\Phi^{\alpha}   &\longrightarrow e^{i\frac{x_{_{\Phi^{\alpha}}}}{\Lambda}a}\Phi^{\alpha},\notag\\
\psi_L &\longrightarrow e^{i\frac{x_{\psi_L}}{\Lambda}a}\psi_L,\notag\\
\psi_R &\longrightarrow e^{i\frac{x_{\psi_R}}{\Lambda}a}\psi_R,\notag\\
S_i &\longrightarrow e^{i\frac{x_{_{S_i}}}{\Lambda}a}S_i,
\end{align}
where $ x_ {\Phi^\alpha} $ and $x_ {\psi ^\alpha_{L,R}} $  are the PQ charges for the Higgs doublets and the SM fermions, respectively.
By keeping the leading order~LO terms in $\Lambda^{-1}$, the Lagrangian~Eq.~\eqref{eq:lagrangian} can be written as~\cite{Georgi:1986df,Brivio:2017ije}: 
\begin{align}
\mathcal{L}_{\text{NLO}}\longrightarrow \mathcal{L}_{\text{NLO}}+\Delta \mathcal{L}_{\text{NLO}}, 
\end{align}
where 
\begin{align}\label{eq:deltaL}
\Delta \mathcal{L}_{\text{NLO}} = \Delta\mathcal{L}_{K^\Phi}+\Delta\mathcal{L}_{K^\psi}+\Delta\mathcal{L}_{\text{Yukawa}}\notag\\
+\Delta\mathcal{L}(F_{\mu\nu})+\Delta\mathcal{L}_{K^S},
\end{align}
with 
{\footnotesize
\begin{align}
 \Delta\mathcal{L}_{K^{\Phi}}=& 
ix_{\Phi^{\alpha}}\frac{\partial^\mu a}{\Lambda}   \left[(D_{\mu}\Phi^\alpha)^{\dagger} \Phi^\alpha
-\Phi^{\alpha\dagger}(D_{\mu}\Phi^\alpha)\right],\notag\\
\Delta\mathcal{L}_{K^\psi}=&
\frac{\partial_\mu 
a}{2\Lambda}\sum_{\psi}(x_{\psi_L}-x_{\psi_R})\bar{\psi}\gamma^{\mu}\gamma^{5}
\psi -(x_{\psi_L}+x_{\psi_R})\bar{\psi}\gamma^{\mu}\psi,\notag\\
\begin{split}
\label{eq30}
 \Delta\mathcal{L}_{\text{Yukawa}}=&    
\frac{ia}{\Lambda}\bar{q}_{Li}\bigg(y^{D\alpha}_{ij}x_{d_j}-x_{q_i}y^{D\alpha}_{
ij}+x_{\Phi^\alpha}y^{D\alpha}_{ij}\bigg) \Phi^{\alpha}d_{Rj}\\
  +&\frac{ia}{\Lambda}\bar{q}_{Li}\bigg(y^{U\alpha}_{ij}x_{u_j}-x_{q_i}y^{U\alpha}_{ij}-x_{\Phi^\alpha}y^{U\alpha}_{ij}\bigg)
  \tilde\Phi^{\alpha}u_{Rj} \notag\\
\Delta\mathcal{L}_{K^S}=&  
ix_{_{S_i}}\frac{\partial^\mu a}{\Lambda}   \left[(D_{\mu}S_i)^{\dagger} S_i
-S^{\dagger}_i (D_{\mu}S_i)\right],
\end{split}
\end{align}}
and $x_{q_i}$, $x_{u_i}$ and  $x_{d_i}$  are the PQ charges 
for the $i$-th 
family of the quark doublet, right-handed up-type and the right-handed down-type, respectively.
From Eq.~\eqref{eq:conditions} we see that $\Delta\mathcal{L}_{\text{Yukawa}}$ is zero, 
this is consistent with the axion shift symmetry which only allows derivative  
couplings to the SM particles. The same is true for all terms without derivatives of the fields.
%%%%%%%%%%%%%%%%%%%%%%%%%%%%%%%%%%%%%%%%%%%%%%%%%%%%%%%%%%%%%%%%%%%%%%%%%%%%%%%%%55
As it is shown in Appendix~\ref{sec:appendix-v-a-couplings} from  $\Delta\mathcal{L}_{K^{\Psi}}$
 we obtain the flavour-violating derivative couplings:
 {\footnotesize
\begin{align}%\label{eq:LKD}
 \Delta\mathcal{L}_{K^D}
  =& -\partial_\mu a
     \bar{d}_{i}\gamma^{\mu}\left(g_{af_if_j}^{V}                         
  +  \gamma^{5}g_{af_if_j}^{A}\right)d_{j},
  \end{align}}
where;
\begin{align}\label{eq:av-couplings0}
g_{ad_id_j}^{V,A}= 
\frac{1}{2f_a c^{\text{eff}}_3}\Delta^{Dij}_{V,A}, 
\end{align}
In this expression we made the substitution $\Lambda = f_a c^{\text{eff}}_3$.  
As shown in appendix~\ref{sec:appendix-v-a-couplings} the axial and vector couplings are:  
\begin{equation}
 \Delta^{Dij}_{V,A}= \Delta^{Dij}_{RR}(d)\pm \Delta^{Dij}_{LL}(q),
\end{equation}
with {\small $\Delta^{Fij}_{LL}(q)= \left(U^D_{L}x_{q}~U_L^{D\dagger}\right)^{ij}$ } and  {\small $\Delta^{Fij}_{RR}(d)= \left(U^D_{R}x_{d}~U_R^{D\dagger}\right)^{ij}_.$} 
 %
%%%%%%%%%%%%%%%%%%%%%%%%%%%%%%%%%%%%%%%%%%%%%%%%%%%%%%%%%%%%
The field redefinitions~\eqref{eq:redefinitions}  induce
a modification of the measure in the functional path integral
whose effects can be determined from the divergence of the 
axial-vector current: 
$J^{PQ5}_{\mu}= 
\sum_{\psi}(x_{\psi_L}-x_{\psi_R})\bar{\psi}\gamma_\mu\gamma^5\psi$~\cite{Bauer:2017ris},
\begin{align}\label{eq:divj5}
\partial^{\mu}J^{PQ5}_{\mu}=&\sum_{\psi}2i m_{\psi}
(x_{\psi_L}-x_{\psi_R})
\bar{\psi}\gamma^5\psi\notag\\
-&\sum_{\psi}(x_{\psi_L}-x_{\psi_R})\frac{\alpha_1    Y^2(\psi)}{2\pi}B_{\mu\nu}\tilde B^{\mu\nu}\notag\\
-&\sum_{\text{ $ SU(2)_L$  doublets}}
\hspace{-0.5cm}
x_{\psi_L}
\frac{\alpha_2}{4\pi}W_{\mu\nu}^a\tilde W^{a\mu\nu}\notag\\
 -&\sum_{ \text{$SU(3)$ triplets}}
 \hspace{-0.5cm}
 (x_{\psi_L}-x_{\psi_R})
 \frac{\alpha_3}{4\pi}G_{\mu\nu}^a\tilde G^{a\mu\nu},
\end{align}
where  the hypercharge  is normalized by
$Q=T_{3L}+Y$.
 The Eq.~\eqref{eq:divj5}
is an on-shell relation; and  the derivative is associated with the momentum of an on-shell axion,
hence, there is internal consistency. 
By replacing this result in $\mathcal{L}_{K^{\psi}}= \frac{\partial^{\mu}a}{2\Lambda}J^{PQ5}_{\mu}= 
-\frac{a}{2\Lambda}\partial^{\mu}J^{PQ5}_{\mu}$
we obtain a modification of the leading order  Wilson 
coefficients~\cite{Salvio:2013iaa}
\begin{align}
c_1& \longrightarrow c_1-\frac{1}{3}\Sigma q+\frac{8}{3}\Sigma u+\frac{2}{3}\Sigma d
-\Sigma \ell+ 2\Sigma e,\notag\\
c_2& \longrightarrow c_2-3\Sigma q-\Sigma \ell,\notag\\
c_3& \longrightarrow c_3-2\Sigma q +\Sigma u+\Sigma d-A_{Q},
\end{align}
where $\Sigma q\equiv x_{q_1}+x_{q_2}+x_{q_3}$ is the sum of the PQ charges of 
the three families, and $A_Q$ is the contribution of the heavy quark to the 
color anomaly which was defined in Eqs.~\eqref{eq:parametrization} 
and~\eqref{eq:parametrization2}.
From these expressions  we obtain for the SM fermions 
{\footnotesize
\begin{align}
 \Delta\mathcal{L}(F_{\mu\nu})=
&\frac{ a}{\Lambda}\frac{\alpha_1}{8\pi}B_{\mu\nu}\tilde B^{\mu\nu}
\left(\frac{1}{3}\Sigma q-\frac{8}{3}\Sigma u-\frac{2}{3}\Sigma d+\Sigma \ell-2\Sigma e\right)\notag\\
+&\frac{ a}{\Lambda}\frac{\alpha_2}{8\pi}W_{\mu\nu}^a\tilde W^{a\mu\nu}
\left(3\Sigma q+\Sigma \ell\right)\notag\\
+&\frac{ a}{\Lambda}\frac{\alpha_3}{8\pi}G_{\mu\nu}^a\tilde G^{a\mu\nu}
\left(2\Sigma q -\Sigma u-\Sigma d+A_{Q}\right).
\end{align}}
We define $c_3^{\text{eff}}=  c_3-2\Sigma q +\Sigma u+\Sigma d-A_{Q}=-N$.
In our case, there are no operators of dimension 5  in the Lagrangian
before redefining the fields, i.e., $c_i=0$.
It is usual to define  $\Lambda  = f_a c_3^{\text{eff}}$
to absorb the factor $c_3^{\text{eff}}$ in the normalization of the PQ charges~\footnote{Notice
that $c_3^{\text{eff}}$ could be negative, however it does not represent a problem since the observables 
always depend on  $\lvert f_a\rvert^2$.}. 
From now on we assume that all the PQ charges are normalized in this way,
so that $x_{\psi}$ stands for  $ x_{\psi}/c_3^{\text{eff}}$ and the effective scale is $f_a$.
For normalized charges $c_{3}^{\text{eff}}=1$, we do not lose generality despite writing the expressions in terms of $ f_a $.

%%%%%%%%%%%%%%%%%%%%%%%%
%%%%%%%%%%%%%%%%%%%%%%%%

%%%%%%%%%%%%%%%%%%%
\section{Naturalness of Yukawa couplings}
\label{sec:IV}
The previous texture analysis guarantees that the number of free parameters in the mass matrices is enough to reproduce the CKM matrix and the quark masses;
as we will show our solutions are flexible enough to set most Yukawa couplings of order 1. 
As shown in the appendices, in order to generate the texture of the mass matrices  with a PQ symmetry, it is necessary to have at least four Higgs doublets.
The chosen PQ charges are enough to generate the texture-zeros;
but it does not guarantee Hermitian mass matrices,   it is true that non-Hermitian mass matrices are the usual ones, however,  in our approach we prefer Hermitian mass matrices to gain some analytical advantages. In order to have self-adjoint matrices we impose the following restrictions on the Yukawa couplings in Eq.~\eqref {eq2}: $ y^{U1}_{31}= y^{U1^*}_{13},y^{U2}_{32}= y^{U2^*}_{23}, y^{D4}_{21}= y^{D4^*}_{12},  y^{D3}_{32}= y^{D3^*}_{23}$,  in addition,  we require that the diagonal elements $y^{U1}_{22},y^{U3}_{33}$ and $y^{D2}_{33}$ must be real numbers.

The  up and down quark  mass matrices  in the interaction basis are:
\begin{align}\label{eq:yij}
M^{U}=\hat{v}_{\alpha} y^{U\alpha}_{ij}= 
\begin{pmatrix}
       0 & 0 & y^{U1}_{13}\hat{v}_1\\\vspace{0.2cm}
       0 & y^{U1}_{22}\hat{v}_1 &  y^{U2}_{23}\hat{v}_2\\
       y^{U1^*}_{13}\hat{v}_1 & y^{U2^*}_{23}\hat{v}_2 & y^{U3}_{33}\hat{v}_3
\end{pmatrix},
\end{align}
\begin{align}\label{eq:yijd}
M^{D}=\hat{v}_{\alpha} y^{D\alpha}_{ij}=
\begin{pmatrix}
 0   & |y^{D4}_{12}|\hat{v}_4 & 0\\
 |y^{D4}_{12}|\hat{v}_4 & 0 & |y^{D3}_{23}|\hat{v}_3\\
 0   & |y^{D3}_{23}|\hat{v}_3 & y^{D2}_{33}\hat{v}_2    
\end{pmatrix},
\end{align}
where we define the expectation values $\hat{v}_i= v_i/\sqrt{2}$.
Here we have implicitly defined the arrays $y^{D\alpha}_{ij}$ which will be needed in the calculation of the FCNC.
 Taking into account the expressions~\eqref{e3.4}, it is possible to establish the following relations between the masses of the up-type quarks and the VEVs
\begin{align}
 \hat{v}_1&=\left(\frac{m_u\, m_c\, m_t}{|y_{13}^{U1}|^2\,y_{22}^{U1}}\right)^{1/3},\label{eq:v1}\\ 
\hat{v}_2&=\sqrt{\frac{(\hat{v}_1\,y_{22}^{U1}-m_u)(\hat{v}_1\,y_{22}^{U1}+m_c)(m_t-\hat{v}_1y_{22}^{U1})}{\hat{v}_1\,
y_ { 22}^{U1}\,|y_{23}^{U2}|^2}},\label{eq:v2a}\\
\hat{v}_3&=\frac{m_u-m_c+m_t-\hat{v}_1\,y_{22}^{U1}}{y_{33}^{U3}}. \label{eq:v3a}
\end{align}
In an identical way for the down sector we can set the following relations:
\begin{align}
\hat{v}_4&=\left(\frac{m_d\, m_s\, m_b}{|y_{12}^{D4}|^2\,(m_d-m_s+m_b)}\right)^{1/2},\label{eq:v4}\\ 
\hat{v}_3&=\sqrt{\frac{(m_s-m_d)(m_d+m_b)(m_b-m_s)}{(m_d-m_s+m_b)\,|y_{23}^{D3}|^2}},\label{eq:v3b}\\
\hat{v}_2&=\frac{m_d-m_s+m_b}{y_{33}^{D2}}. \label{eq:v2b}
\end{align}
By using current quark masses  at the $Z$ pole (Table~\ref{tab:eqF3}), i.e., $m_u=1.27$~MeV, $m_c=0.633$~GeV and
$m_t=171.3$~GeV, from Eq.~\eqref{eq:v1} we find the following approximate values for the vacuum 
expectation in terms of  the masses and the Yukawas:
\begin{align}
\hat{v}_1 y_{22}^{U1}\sim \left\lvert\frac{y_{22}^{U1}}{y_{13}^{U1}}\right\rvert^{2/3} (m_u m_c m_t)^{1/3}
=\left\lvert\frac{y_{22}^{U1}}{y_{13}^{U1}}\right\rvert^{2/3} 0.516\,\text{GeV}.    
\end{align}
From the bottom current mass at the $Z$ pole  we can obtain $\hat{v}_2$ by using the Eq.~(\ref{eq:v2b})
\begin{equation}
\hat{v}_2\sim \frac{m_b}{y_{33}^{D2}}=\frac{2.91\, \text{GeV}}{y_{33}^{D2}}.
\label{eq18}
\end{equation}
Using the constraint~\eqref{eq5} and the numerical inputs in Table~\ref{tab:eqF3} in Appendix~\ref{sec:mat-diag}, we can establish the more restrictive condition~$ m_u\ll y_{22}^{U1}\, \hat{v}_1 \ll m_t$. The consistency between the  equations~\eqref{eq:v2a} and ~\eqref{eq18} requires the following relation
\begin{align}
 \left\lvert \frac{y^{U2}_{23}}{y^{D2}_{33}} \right\rvert
 =\sqrt{ \frac{\left( m_c+\hat{v}_1y^{U1}_{22}\right) m_t}{m_b^2}}\sim 6.9, 
\end{align}
where we are assuming   that  $\hat{v}_1y^{U1}_{22} \sim2.7\, m_c$~(see Table~\ref{tab:eqF3}). 
Under similar assumptions it is also possible to get $\hat{v}_3$
from the equation~\eqref{eq:v3a}
\begin{align}\label{eq:v3}
\hat{v}_3\sim \frac{m_t}{y_{33}^{U3}}. 
\end{align}
The consistency of this result with the value for $\hat{v}_3$ in Eq.~\eqref{eq:v3b} implies
\begin{align}
\left\lvert \frac{y^{D3}_{23}}{y^{U3}_{33}} \right\rvert
 =\sqrt{\frac{m_s m_b}{m_t^2}}= 2.4\times 10^{-3},
\end{align}
where, in this case,  we took  $m_s =56$~MeV at the $Z$ pole.
Due to Eq.~\eqref{eq:v3} all the  Yukawa couplings have  a strong dependency on $y^{U3}_{33}$ since $\hat{v}_3$ is 
the leading term in $v=\sqrt{(v_1^2+v_2^2+v_3^2+v_4^2)}$. 
So, by setting  various Yukawa couplings close to 1 (except 
$y^{U2}_{23}$, $
y^{D3}_{23}$  and  $y_{13}^{U1}$) we obtain: 
\begin{align}\label{eq:vevs}
\hat{v}_1 =&  1.71 \,\text{GeV},\hspace{0.2cm}
\hat{v}_2 =  2.91 \, \text{GeV}, \hspace{0.2cm}
\hat{v}_3 =  174.085 \, \text{GeV}.
%\hspace{0.2 cm}
\end{align}
Finally, we can obtain $\hat{v}_4$ from Eq.~\eqref{eq:v4}
\begin{align}
\hat{v}_4\sim \frac{\sqrt{m_d\, m_s}}{\lvert y^{D4}_{12}\rvert}. 
\end{align}
By setting  $y_{33}^{U3}\sim 0.983818$ it is possible to adjust  $y_{12}^{D4}\sim 1$ through the relation
 $(v_1^2+v_2^2+v_3^2+v_4^2)= (246.24\,\text{GeV})^2=v^2$,
which  for $m_d= 3.15$~MeV implies   
\begin{align}\label{eq:vevs2}
\hat{v}_4 = 13.3  \,\text{MeV}.    
\end{align}
%%%%%%%%%%%%%%%%%%%%%%
We will adjust the scalar potential $V(\Phi,S_1,S_2)$ so that, at the minimum, the VEVs of the scalar doublets are precisely those required to generate the SM quark masses.  We also propose rotation matrices to implement the Georgi-Nanopoulus formalism for an arbitrary number of scalar doublets.  
%%%%%%%%%%%%%%%%%%%%%%%%%%%%%%%%%%%%%%%%%%%%%%%%%%%%%
\section{Low energy  constraints}
\label{sec:constraints}
Since our model has non-universal PQ charges, in addition to the usual constraints for the axion-photon coupling, a tree level analysis of the Flavor Changing neutral currents is needed. 
As it is mentioned in reference~\cite{DiLuzio:2020wdo} the strongest bounds on flavor violating axion couplings to quarks come from meson decays into final states containing invisible particles. Currently, the $K^{\pm}\rightarrow \pi^{\pm}a$ decays provide the tightest limits (E949 and E787 Experiments) for the axion mass~\cite{DiLuzio:2020wdo}. Other important restrictions apply on axion-photon couplings~\cite{DiLuzio:2020wdo} but require lepton couplings which we are not considering in this work, any way,  in our case these bounds do not represent the strongest constraints~\cite{DiLuzio:2020wdo}.
As shown in reference~\cite{DiLuzio:2020wdo} for the decays $K^{\pm}\rightarrow \pi^{\pm}a$  and  
$B\rightarrow K^{*}a$  the tree level FCNC come from  the term $\Delta\mathcal{L}_{K^{\psi}}$ in the Lagrangian~\eqref{eq:deltaL}. 
In our approach, we assume that these terms are absent in the original Lagrangian, i.e.,  $c_i=0$, so these terms 
come from the redefinition of the  fields~\eqref{eq:redefinitions} and are therefore proportional to the PQ charges. 
%%\begin{widetext}
%\scalebox{0.8}{
\begin{figure}%[h!]
%\vspace{-96pt}
\begin{center}
\centering 
\begin{tabular}{cc}
\includegraphics[scale=0.3]{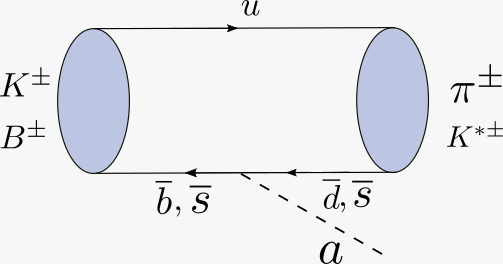}% &  \includegraphics[scale=0.3]{AI.pdf}\\   
\end{tabular}
\end{center}
\caption{ Tree level diagram contribution to the FCNC processes 
$K^{\pm}\rightarrow \pi^{\pm}a$  and $B^{\pm}\rightarrow K^{*\pm}a$.}
\label{fig:ktopi}	
\end{figure} 
In Appendix~\ref{sec:appendix-v-a-couplings}, it is shown that the decay widths
of pseudoscalar $K^{\pm}$(B)  mesons into an axion and a  charged pion~(vector $K^*$) are given by
 \begin{align}
 \Gamma(K^\pm\rightarrow \pi^\pm a) =& \frac{m_K^3}{16\pi}\left(1-\frac{m_\pi^2}{m_K^2}\right)^2\lambda_{K\pi a}^{1/2} f^2_0(m_a^2)
 \lvert g_{ads}^V \rvert^2,\notag\\
 \Gamma(B\rightarrow K^{*}a) =& \frac{m_B^3}{16\pi}
 \lambda_{BK^{*}a}^{3/2} A_{0}^2(m^2_a)\lvert g_{asb}^A\rvert^2,  
  \end{align}
where $\lambda_{Mm a}= \left(1-\frac{(m_a+m)^2}{M^2}\right)\left(1-\frac{(m_a-m)^2}{M^2}\right)$
and 
\begin{align}\label{eq:av-couplings}
g_{ad_id_j}^{V,A}= 
\frac{1}{2f_a c^{\text{eff}}_3}\Delta^{Dij}_{V,A}, 
\end{align}
where:
\begin{equation}
 \Delta^{Dij}_{V,A}= \Delta^{Dij}_{RR}(d)\pm \Delta^{Dij}_{LL}(q),
\end{equation}
 \noindent with {\small $\Delta^{Fij}_{LL}(q)= \left(U^D_{L}x_{q}~U_L^{D\dagger}\right)^{ij}$ } and  {\small $\Delta^{Fij}_{RR}(d)= \left(U^D_{R}x_{d}~U_R^{D\dagger}\right)^{ij}_.$} 
In the Eq.~\eqref{eq:av-couplings} we normalize the charges with $c^{\text{eff}}_3$ as it is explained in the last paragraph of section~\ref{sec:effectiveL}. 
\begin{table}
 \begingroup
\renewcommand*{\arraystretch}{1.5} 
 \begin{equation}
% {|l|l|}
 \begin{array}{lc}
 \hline 
 \hline 
 \hspace{0.5cm} \text{Collaboration} & \text{Upper bound} \\
 \hline
 \text{E949+E787~\cite{Adler:2008zza,Artamonov:2008qb}} 
            &\mathcal{B}\left(K^+\rightarrow\pi^+      a\right)<0.73\times 10^{-10}\\
\text{CLEO~\cite{Ammar:2001gi}}                 
            &\mathcal{B}\left(B^\pm\rightarrow \pi^\pm a\right)<4.9\times 10^{-5}\\
 \text{CLEO~\cite{Ammar:2001gi}}
            &\mathcal{B}\left(B^\pm\rightarrow K^\pm   a\right)<4.9\times 10^{-5} \\
 \text{BELLE~\cite{Lutz:2013ftz} }
            &\mathcal{B}\left(B^\pm\rightarrow \rho^{\pm} a\right)<21.3\times 10^{-5} \\                        
 \text{BELLE~\cite{Lutz:2013ftz} }
            &\mathcal{B}\left(B^\pm\rightarrow K^{*\pm}   a\right)<4.0\times 10^{-5} \\            
\hline
\hline 
 \end{array}
\end{equation}
  \caption{These inequalities come from the window for new physics in the branching ratio uncertainty of the meson decay in a pair $\bar{\nu}\nu$.}
 \label{tab:fcnc} 
 \endgroup  
\end{table}
\begin{figure}
\begin{center}
\centering 
\begin{tabular}{cc}
 \includegraphics[scale=0.35]{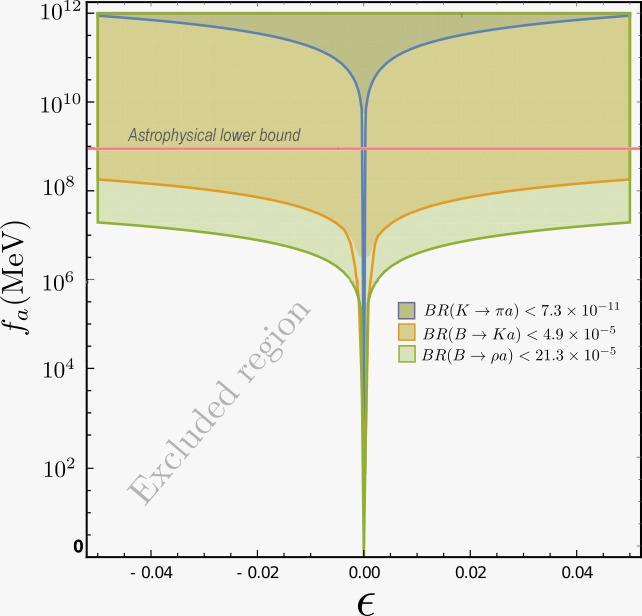}  
 \end{tabular}
\end{center}
%\vspace{-4cm}
\caption{Allowed regions for semileptonic meson decays. We use the 
relation~\eqref{eq1x} between the axion mass and the decay constant $f_a$.}
\label{fig:fa_vs_epsilon}	
\end{figure}
For $m_a \ll$ 1~MeV, the  form factor is
$f_0(m_a^2)\approx 1$~\cite{Carrasco:2016kpy}
 for the decay $K^{\pm}\rightarrow\pi^{\pm}a$.  
On the other side, from reference~\cite{Ball:2004ye} we obtain:
$f_0(m_a^2)\approx 0.33$
for $B^{\pm}\rightarrow K^{\pm}a$,
$f_0(m_a^2)\approx 0.258$ 
for $B^{\pm}\rightarrow\pi^{\pm}a$, 
 and for decays with a vector meson in the final state  $A_0(m_a^2)\approx 0.374$ for $B^{\pm}\rightarrow K^{*\pm}a$. 
The constraints on the axion couplings and the decay constant $f_a$  can be obtained from rare semileptonic meson decays $M\rightarrow m \bar{\nu}\nu$, where $M$ stands for $K^{\pm},B^{\pm}$  and $m=\pi^{\pm},K^{\pm},K^{*},\rho$. These constraints are summarized in Table~\ref{tab:fcnc}.
Figure~\ref{fig:fa_vs_epsilon} shows the decay  constant $f_a$ as a function of $\epsilon$.
For our PQ charges, the FCNC from the processes $B^\pm\rightarrow \pi^{\pm}a$ and $B^\pm\rightarrow K^{*\pm}a$ are strongly suppressed, in such a way that these constraints are satisfied trivially, hence their allowed regions are not shown in Figure~\ref{fig:fa_vs_epsilon}.

In general, it is not guaranteed that the eigenstates of mass correspond to the states obtained from the Georgi Rotation, as it is argued in the reference~\cite{Das:2019yad} it is only necessary that the state corresponding to the Higgs of the SM coincides with one of the mass eigenstates of the neutral scalars to obtain an alignment that allows us applying the results of the formalism of Georgi~\cite{Georgi:1978ri}. In our case, we have numerically verified the alignment criteria in reference~\cite{Das:2019yad}.
The origin of the alignment in our model is a consequence of the large suppression of the VEVs of the scalar doublets
$v_i$, with $i=1,2,4$, respect to $v_3$, the VEV of $\Phi_3$.
To some extent, this alignment avoids FCNC involving the SM Higgs boson; however, after alignment, there are other sources of FCNC associated with the additional scalar doublets, which is not possible to avoid by any means. 

New sources of FCNC come from the Higgs sector, as can be seen in Eq.~\eqref{eq:yuk} in appendix~\ref{sec:georgiB},  where the term $  -\bar{d}_L^{i}   H_\beta^0    Y_{ij}^{D\beta}     d^{j}_{R}
              -\bar{u}_L^{i}   H_\beta^{0*} Y_{ij}^{U\beta}     u^{j}_{R}$ has FCNC for $\beta=2,3,4$, however, for $\beta=1$   $Y_{ij}^{U1}$ is diagonal,  $H_1^{0*}$ corresponds to the SM Higgs field, hence, 
there are no terms with flavor-changing neutral currents involving the SM Higgs. 
For $\beta=2,3,4$ the decay $B\rightarrow K^*H^{\beta}$ with a neutral scalar in the final state has no phase space, however,
the FCNC process $B\rightarrow K^*H^{\beta}\rightarrow K^* \ell^-\ell^+$ , where the scalar is an intermediate boson, is possible, however, in this case, the scalar width is suppressed by a factor $1/M_{\beta}^4$ (for $\beta>1$ the masses are above 1TeV)
and therefore this observable does not represent the strongest constraint. This justifies why in the literature the width of the FCNC process $\pi\pm\rightarrow K^\pm a $ (Eq. 39) represents the strongest constraint for a light axion.    
The PDG 2022~\cite{Workman:2022ynf}, set  mass limits for heavy neutral Higgs bosons in the MSSM (which is a usual benchmark model for models with additional Higgs doublets)  $M_{2}>389$~GeV for $\tan \beta = 10$. The constraints are stronger for larger $\tan \beta$; in our model, the $\tan \beta$ values are of order one so that  in all the cases the scalar masses of our model are above these lower limits. 

From astrophysical considerations are the bounds from black holes superradiance and the SN 1987A bound on the neutron electric dipole moment, which can be combined  in such a way that they constrain the axion decay constant in the range~\cite{DiLuzio:2020wdo} (see Figure~\ref{fig:fa_vs_epsilon}) :
$
    0.8\times10^{6}\text{GeV}\leq f_a\leq 2.8\times 10^{17}\text{GeV}
$.

%Although an interpretation of the  XENON1T anomaly \cite{XENON:2020rca}  in terms of solar  axions is in strong tension with astrophysical observations of stellar evolution~\cite{DiLuzio:2020jjp}, a recent analysis showed that inclusion of the inverse Primekoff process has a significant impact on the parameter region preferred by the XENON1T data, significantly reducing the tension with the stellar
%cooling bounds. Even though the tension remains this analysis  shows that this subject requires a critical review~\cite{Gao:2020wer} before completely ruling out this possibility.
%In this vein, we showed that by choosing conveniently the parameters,  our model can adjust 
%the axion mass required to explain the anomaly recently reported by 
%XENON1T~\cite{XENON:2020rca}. This is a result of the flexibility of our approach, which 
%is an important feature of any model useful for approaching phenomenological studies.
%By choosing conveniently the parameters,  it is possible to adjust  the recently
%reported XENON1T anomaly.
%%%%%%%%%%%%%%%%%%%%%%%%%%%%%%%%%%%%%%%%%%%%%%%%%%%%%%%%%%%%%%%%%%%%%%%%%
%%%%%%%%%%%%%%%%%%%%%%%%%%%%%%%%%%%%%%%%%%%%%%%%%%%%%%%%%%%%%%%%%%%%%%%%%
\section{Summary and conclusions }
\label{sec:conclusions}
 In this work we have proposed a PQ symmetry that gives rise to quark mass 
matrices 
with five texture-zeros.
This texture~\eqref{5.1y}  can adjust in a 
non-trivial way the six masses of the quarks and the three CKM mixing angles  
and the CP violating phase. The Hermitian quark mass matrices, up-type $M^U$ and  down-type $M^D$, have 18 
free parameters,  six of them are phases and 12 are real parameters. As it is 
well known in the literature, three of these real parameters can be made equal to 
zero through a WBT  without any physical 
consequence~\cite{Fritzsch:1999ee,Branco:1988iq,Branco:1999nb}.  Five of these 
phases can be reabsorbed in the fermion 
fields~\cite{Kobayashi:1973fv,Maiani:1975in} in such a way that we end with nine 
real parameters and one phase to explain the six quark masses, the three mixing 
angles, and the CP-violating phase, achieving parity between the number of free parameters 
and experimental measurements.

By imposing two texture zeros~(in addition to the 
three zeros obtained from the~ WBT)  there are more experimental constraints 
than free parameters, this feature eliminates a large number of possible 
textures for the mass matrices.
In Appendix~\ref{sec:demostracion} we showed that in order to generate  the  texture, Eq.~\eqref{5.1y}, through
a PQ symmetry, at least four Higgs doublets  are required. 
In Eq.~\eqref{eq:parametrization} we proposed a general parametrization for the PQ 
charges which is consistent with the texture. 

Since many observables are 
proportional to the PQ charges normalized by the QCD anomaly, we included 
into the particle content  a heavy quark singlet under 
the SM gauge electroweak gauge group $SU_L(2)\times U_Y(1)$  but   with chiral charges under the PQ symmetry.
The PQ charges of this heavy quark are responsible for maintaining $N\neq 0$, while we make the PQ charges of the SM quarks arbitrarily close to zero.

To generate the texture zeros of the mass matrices and simultaneously to solve the strong CP 
problem  it is necessary to keep $\epsilon$ and $N$ different from zero in 
Eq.~\eqref{eq:parametrization}.  
In our case, the FCNC observables do not depend 
on the parameters $\alpha $ and $\hat{s}_1$ (see Table~\ref{tab:pcontent2} for definitions),
hence, the axion decay constant $f_a$  (or the axion mass $m_a$) and  $\epsilon$ were the only relevant parameters in our analysis.

In order to write down the quark mass matrices in the proper basis, in Appendix~\ref{sec:georgiB} we generalize the Georgi 
rotation in the two Higgs doublet formalism
to rotate an arbitrary number of Higgs doublets to a basis where only one Higgs doublet acquires a vacuum expectation value. 

By defining almost all the  Yukawas close to 1,  it was possible to determine 
the vacuum expectation values of the Higgs doublets from the experimental value 
of the quark masses and the CKM mixing matrix, this choice obeys the criteria of 
naturalness and is very convenient to understand the origin of the mass 
hierarchies in the SM. 

Since in our model the PQ charges are non-universal there 
are FCNC at the tree level.  We calculated  the tree level FCNC couplings 
from the effective interaction Lagrangian  between the kinetic term of the 
quarks and the axion, these couplings are well known in the 
literature~\cite{DiLuzio:2020wdo}.
In Appendix~\ref{sec:appendix-v-a-couplings} we calculated the decay width for 
the decay of a pseudoscalar meson into a pseudoscalar~(or vector) meson plus an 
axion. This result let us determine the region of the parameter space allowed by 
the experimental constraints Fig.~\ref{fig:fa_vs_epsilon}.

Our model is flexible enough to accommodate possible experimental anomalies, while simultaneously 
is a useful approach to answer several issues in flavor physics. 
For future work, it is necessary to extend the PQ symmetry to leptons. Although 
it is true that in the literature there are textures that can adjust the 
parameters of the lepton 
sector~\cite{Fritzsch:2009sm,Fritzsch:2015gxa,Fritzsch:2015haa,Fritzsch:2016xmb,Hollik:2014jda,Benavides:2020pjx}, these textures are different from those used 
with quarks~\cite{Fritzsch:1999ee}, however, as will be shown elsewhere, it is possible to make the adjustment
without additional Higgs doublets.
%It is possible to use textures with a 
%smaller number of zeros. This election requires fewer Higgs doublets but in this 
%case, it is difficult to keep the Yukawa couplings close to 1, therefore, it is 
%not possible to obtain an insight into the origin of the SM hierarchies.
%%%%%%%%%%%%%%%%%%%%%%%%%%%%%%%%%%%%%%%%%%%%%%%%%%%%%%%%%%%%%%%%%

%%%%%%%%%%%%%%%%%%%%%%%%%%%%%%%%%%%%%%%%%%%%%%%%%%%%%%%%%%%%%%%%%
\section*{Acknowledgments} 
We thank Financial support from ``Patrimonio Aut\'onomo Fondo Nacional de Financiamiento para la Ciencia, la Tecnolog\'ia
y la Innovaci\'on, Francisco Jos\'e de Caldas''.
This research was partly supported by the ``Vicerrectoría de Investigaciones e Interacción Social VIIS de la Universidad de Nariño'',  project numbers 024, 160, 1928 and 2172.

\appendix
%%%%%%%%%%%%%%%%%%%%%%%%%%%%%%%%%%%%%%%%%%%%%%%%%%%%%%%%%%%%%%%%%
%%%%%%%%%%%%%%%%%%%%%%%%%%%%%%%%%%%%%%%%%%%%%%%%%%%%%%%%%%%%%%%%%
\section{The minimal content of  Higgs doublets}
\label{sec:demostracion}
% \subsection*{Demostración}%%%%%%%%%%%%%%%%%%%%%%%%
The texture~\eqref{MuMd} can be obtained from  a Peccei-Quinn  $U(1)_{PQ}$ symmetry, incorporating in the model a minimum of 4 Higgs doublets with charges $x_\phi$.\begin{equation}\label{MuMd}
M^{U}=\left(\begin{array}{ccc}
       0 & 0 & x\\
       0 & x & x\\
       x & x & x
      \end{array}\right),
      \hspace{1cm}
      M^{D}=\left(\begin{array}{ccc}
       0 & x & 0\\
       x & 0 & x\\
       0 & x & x
      \end{array}\right).
\end{equation}
The idea of the demonstration is:
we first observe that in terms of the  charges $x_{\psi}$, each entry allowed in the array $M^{U}$ must satisfy the relation:
\begin{equation}
S_{ij}^{U}=-x_{q_{i}}+x_{u_{j}}-x_{\phi_{\alpha}}=0,
\label{Sij}
\end{equation}
where $x_{\phi_\alpha}$  represents the PQ chargue of the $\alpha$th Higgs that satisfies the equality in Eq.~\eqref{Sij}.
By assuming two quarks doublets $q_{L_i}$ and $q_{L_j}$ with identical PQ charges  $x_q$ 
and requiring  $S_{ik}^{U}=-x_{q_{i}}+x_{u_{k}}-x_{\phi_{\alpha}}=0 $ for any $k=1,2,3$, we also have  $S_{jk}^{U}=-x_{q_{j}}+x_{u_{k}}-x_{\phi_{\alpha}}=0$, for the same $k$'s and the Higgs doublet  
$\phi_{\alpha}$~(since $x_{q_i}=x_{q_j}$). This would lead to having two rows in the matrix $M^U$ with an equivalent structure, that is to say,  the allowed and 
forbiden terms are the same, which contradicts the structure of the matrix. Similarly, if two fields $u_{Ri}$, $u_{Rj}$ with $i\ne j$, had equal charges, it would lead to an array $M^U$ with a similar structure in two columns, which is not present in~\eqref{MuMd}; the same applies to the matrix $M^D$, thus:

\begin{align}\label{eq:xinexj}
  x_{q_i}\ne x_{q_j},\hspace{0.3cm} 
  x_{u_i}\ne x_{u_j},\hspace{0.3cm}
  x_{d_i}\ne x_{d_j},\hspace{0.3cm}
   \text{with}\hspace{0.2cm} i,j=1,2,3.
\end{align}  

From these inequalities and noting that in the third column in $M^U$ all terms are allowed, we can conclude that at least three Higgs doublets are required to reproduce the texture-zeros of the matrix $M^U$.
Now it is necessary to settle  if three Higgs doublets are enough to simultaneously reproduce the matrix $M^U$ and $M^D$ in~\ref{MuMd}. 
The   third column in $M_{u}$ implies the relations
\begin{align}\label{eq:s3}
S^{U}_{i,3}=-x_{q_{i}}+x_{u_{3}}-x_{\phi_{\alpha}}=0,\hspace{0.5cm} \text{for each}\hspace{0.5cm} i=1,2,3,
\end{align} 

then $x_{q_{1}}=x_{u_{3}}-x_{\phi_{\alpha}}=0$  and $x_{q_{2}}=x_{u_{3}}-x_{\phi_{\alpha}}=0$.
Since $x_{q_1}\ne x_{q_2}$ these equations can not be simultaneously valid for the same 
$x_{\phi_\alpha}$.
The same is true for any pair  $x_{q_{i}}$,  $x_{q_{j}}$ with $i\neq j$,
hence, the equalities Eq.~(\ref{eq:s3}) require a minimum of three higgs doublets to reproduce  the texture of $M^U$.
The next step is to determine if the three  chosen Higgs doublets for 
$M^{U}$ are enough to generate the texture of $M^{D}$.
For three Higgs doublets the texture~\eqref{MuMd} requires $7\times 3=21$ inequalities 
associated with the forbidden entries, i.e., 
\begin{align}
S_{ij}^D =&-x_{q_{i}}+x_{d_{j}}+x_{\phi_\alpha}\neq 0\notag\\
S_{ij}^{U} =&-x_{q_{i}}+x_{u_{j}}- x_{\phi_\alpha}\neq 0, \hspace{0.5cm}\alpha=1,2,3.
\label{Sijneq}
\end{align}
Now, without loss of generality, we can take the charge of the singlet $x_{u_3}=0$, and from the equations~\eqref {eq:s3} for the couplings of $u_ {R_3}$ we can identify the charges of the doublets $q_L $ with the charges of the three Higgs fields, such that: $x_ {q_i} = - x _ {\phi_i}$. With this result we can put together the equations~\eqref{Sij} and the inequalities~\eqref{Sijneq}, in such a way that the texture of the matrix 
$ M^{D} $ can be written-down as \ref{SumXd}:

\begin{strip}
\begin{equation}\label{SumXd}
S^D =\left(\begin{array}{ccc}
       x_{\phi_1}+x_{d_1}+(x_\phi)_{11}\neq 0 & x_{\phi_1}+x_{d_2}+(x_\phi)_{12}= 0 & x_{\phi_1}+x_{d_3}+(x_\phi)_{13}\neq 0\\
       x_{\phi_2}+x_{d_1}+(x_\phi)_{21}= 0 &x_{\phi_2}+x_{d_2}+(x_\phi)_{22}\neq 0 & x_{\phi_2}+x_{d_3}+(x_\phi)_{23}= 0\\
       x_{\phi_3}+x_{d_1}+(x_\phi)_{31}\neq 0 & x_{\phi_3}+x_{d_2}+(x_\phi)_{32}= 0 & x_{\phi_3}+x_{d_3}+(x_\phi)_{33}= 0
\end{array}\right),
\end{equation}
\end{strip}
where the inequalities must be satisfied by any  $(x_\phi)_{ij}=x_{\phi_k}$, with  $k=1,2,3$.
For the equalities, it is enough if at least one  $x_{\phi_i}$ satisfies them. In Eq.~\eqref{cas} we analyze each entry of $S_{21}^D$ and we obtain the following options for $(x_\phi)_{21}$:
\begin{align}
(x_\phi)_{21}
=
\begin{cases}
x_{\phi_1} & \rightarrow  S^{D}_{11}=0~ \text{if}~ (x_{\phi})_{11}=x_{\phi_2}~(\text{must be}\ne 0) \\
x_{\phi_2} &  \text{is a consistent solution}  \\
x_{\phi_3} & \rightarrow  S^{D}_{13}=0~ \text{if}~ (x_{\phi})_{13}=x_{\phi_2}~(\text{must be}\ne 0), 
\end{cases} 
\label{cas}
\end{align}

%Siguiendo el analisis presentado,escoger
By the same way, the choice 
$(x_\phi)_{23}=x_{\phi_1}$ in $S^D_{23}$ 
%viola la desigualdad
is not consistent with the inequality $S^D_{13}$, 
%y escoger
and  the choice 
$(x_\phi)_{23}= x_{\phi_2}$
due to $S^D_{21}$,
implies  $x_{d_3}=x_{d_1}$,
%que $x_{d_3}=x_{d_1}$, 
which is forbidden by Eq.~\eqref{eq:xinexj}, therefore the only option is
$(x_\phi)_{23}=x_{\phi_3}$.
The proposed analysis allows defining in a unambiguous way the fields $(x_\phi)_{ij}$  in the equalities. Proceeding in an identical way for the remaining ones, we get:
 \begin{align}
 (x_\phi)_{21}=&x_{\phi_2}, \hspace{0.3cm} (x_\phi)_{23}=x_{\phi_3}, \hspace{0.3cm}(x_\phi)_{12}=x_{\phi_3},\notag\\
 (x_\phi)_{32}=&x_{\phi_1}, \hspace{0.3cm} (x_\phi)_{33}=x_{\phi_2},
  \end{align}
By replacing these expressions in~\eqref{SumXd} $S^D$ reduces to \ref{SumXd2}:

\begin{strip}
\begin{equation}\label{SumXd2}
S^{D} =\left(\begin{array}{ccc}
       x_{\phi_1}+x_{d_1}+(x_\phi)_{11}\neq 0 & x_{\phi_1}+x_{d_2}+x_{\phi_{3}}= 0 & x_{\phi_1}+x_{d_3}+(x_\phi)_{13}\neq 0\\
       2x_{\phi_2}+x_{d_1}= 0 & x_{\phi_2}+x_{d_2}+(x_\phi)_{22}\neq 0 & x_{\phi_2}+x_{d_3}+x_{\phi_{3}}= 0\\
       x_{\phi_3}+x_{d_1}+(x_\phi)_{31}\neq 0 & x_{\phi_3}+x_{d_2}+x_{\phi_{1}}= 0 & x_{\phi_3}+x_{d_3}+x_{\phi_{2}}= 0
\end{array}\right).
\end{equation}
\end{strip}

From this expression we obtain the relation:

\begin{equation}
    S^D_{21}-S^D_{11}=2x_{\phi_2}-x_{\phi_1}-(x_\phi)_{11}\neq 0,
\end{equation}
\noindent since this must be true for all $(x_\phi)_{11}=x_{\phi_i}$, for $i=1$ we get: 

\begin{equation}
  2x_{\phi_2}-x_{\phi_3}-x_{\phi_{1}}\neq 0.
 \label{restricciones}
\end{equation}

We will use equation (\ref{restricciones}) shortly. By carrying out the same analysis for $S^{U}$~(using the same conventions $x_{u_3}=0$ and $-x_{q_i}=x_{\phi_i}$ ) there are two options for this matrix, as seen in \ref{Su}:
{\small
\begin{equation}\label{Su}
S^U_{A \choose B} =\left(\begin{array}{ccc}
      x_{\phi_1}+x_{u_1}-(x_{\phi})_{11}\neq0  & x_{\phi_1}+x_{u_2}-(x_{\phi})_{12}\neq 0  & 0\\
       x_{\phi_2}+x_{u_1}-(x_{\phi})_{21}\neq 0 & x_{\phi_2}+x_{u_2}-x_{\phi_{1 \choose 3}}=0  & 0\\
       x_{\phi_3}+x_{u_1}-x_{\phi_{1 \choose 2}}=0 & x_{\phi_3}+x_{u_2}-x_{\phi_{2\choose 1}}=0 & 0
\end{array}\right),
\end{equation}}

where subscript $2\choose 1$ indicates that either of the two values $x_{\phi_1}$ or $x_{\phi_2}$ are possible.
The subscript $A\choose B$  means that all the 
up~(down) options must be replaced simultaneously,  mixing between up and down 
options must be avoided.
From this matrix, i.e., $S^U_{A}$, we obtain  $(S^U_{A})_{22}-S^U_{A})_{32}=2x_{\phi_2}-x_{\phi_3}-x_{\phi_{1}}=0$, 
which is forbidden by~\eqref{restricciones},
then, the option $S^U_{A}$ is not possible.  
For the option $S^U_{B}$ we have 
\begin{equation}
(S^U_{B})_{22}-(S^U_{B})_{32}=-2x_{\phi_3}+x_{\phi_2}+x_{\phi_{1}}=0, 
\label{suneq}
\end{equation}
but, $(S^U_{B})_{11}-(S^U_{B})_{31}=-2x_{\phi_3}+x_{\phi_2}+x_{\phi_{1}}\neq0$ (where we took   $(x_{\phi})_{11}=x_{\phi_3}$ in $S^U_B$)
that violates the inequality \eqref{suneq}, therefore it is not possible to build the texture~\eqref{MuMd} with just three Higgs doublets.
By adding a Higgs doublet, infinite solutions are presented thus demonstrating that a minimum of four Higgs doublets are required to reproduce the texture~\eqref{MuMd}.

%%%%%%%%%%%%%%%%%%%%%%%%%%%%%%%%%%%%%%%%%%%%%%%%%%%%%%%%%%%%%%%%%%%%%%%%%%%%%%
%%%%%%%%%%%%%%%%%%%%%%%%%%%%%%%%%%%%%%%%%%%%%%%%%%%%%%%%%%%%%%%%%%%%%%%%%%%%%%
\section{The mass operator matrices} 
\label{sec:georgiB}
The most general Lagrangian for the interaction of four Higgs  doublets $\Phi_\alpha$  with the quarks of the SM is given by
\begin{align}\label{eq:yukawa}
\mathcal{L}= 
 -\bar{q}_L^{\prime i} \Phi_\alpha y_{ij}^{D\alpha}d^{\prime j}_{R}
 -\bar{q}_L^{\prime i} \tilde\Phi_\alpha y_{ij}^{U\alpha}u^{\prime j}_{R}+\text{h.c},
\end{align}
where a sum is assumed on repeated indices.
Here  $i,j$ run over $1,2,3$ and $\alpha$ over $1,2,3,4$. 
The Higgs  boson doublet fields are parameterized as follows:
\begin{align}\label{eq:higgs}
\Phi_\alpha = 
\begin{pmatrix}
\phi_\alpha^{+}\\
\frac{v_\alpha+h_\alpha+i\eta_\alpha}{\sqrt{2}}
\end{pmatrix},
\hspace{1cm}
\tilde\Phi_\alpha=i\sigma_2 \Phi_\alpha^{*}.
\end{align}
In a similar way as in the two Higgs doublet  model~\cite{Cardozo:2020uol} 
we  rotate the Higgs fields  to the (generalized) Georgi basis, i.e.,
\begin{align}\label{eq:rab}
\begin{pmatrix}
H_1 \\
H_2 \\
H_3 \\
H_4 
\end{pmatrix}
= 
R_1(\beta_1)R_2(\beta_2)R_3(\beta_3)
\begin{pmatrix}
\Phi_1\\
\Phi_2\\
\Phi_3\\
\Phi_4
\end{pmatrix}=: H_{\beta}
\equiv R_{\beta\alpha}\Phi_\alpha , 
\end{align}
where the orthogonal matrices
\begin{subequations}
\label{eqC4}
\begin{equation}
R_1(\beta_1)=
\begin{pmatrix}
 \cos \beta_1 & \sin \beta_1 & 0 & 0\\
-\sin \beta_1 & \cos \beta_1 & 0 & 0\\
      0     &   0        & 1 & 0\\    
      0     &   0        & 0 & 1          
\end{pmatrix},
\end{equation}
\begin{equation}
R_2(\beta_2)=
\begin{pmatrix}
      1     &   0        & 0                & 0\\          
      0     & \cos \beta_2 & \sin \beta_2   & 0 \\
      0     &-\sin \beta_2 & \cos \beta_2   & 0 \\
      0     &   0          & 0              & 1\\    
\end{pmatrix},
\end{equation}
\begin{equation}
R_3(\beta_3)=
\begin{pmatrix}
      1     &  0              &   0          & 0              \\
      0     &  1              &   0          & 0              \\ 
      0     &  0              & \cos \beta_3 & \sin \beta_3   \\
      0     &  0              &-\sin \beta_3 & \cos \beta_3   \\
\end{pmatrix},
\end{equation}
\end{subequations}
where 
$\tan \beta_1 =\frac{\sqrt{v_2^2+v_3^2+v_4^2}}{v_1}$,
$\tan \beta_2 =\frac{\sqrt{v_3^2+v_4^2}}{v_2}$ and 
$\tan \beta_3 =\frac{v_4}{v_3}$. 
In these expressions $H_\beta= (H_\beta^+,(H_\beta^0+iH_{\beta}^{\text{odd}})/\sqrt{2})^T$.
This basis is chosen in such a way that only the neutral component of $ H_1 $ 
acquires
a vacuum expectation value
\begin{align}
&\langle H_1^0 \rangle =\sqrt{v_1^2+v_2^2+v_3^2+v_4^2}\equiv v%\\
,\hspace{0.5cm}\notag\\
&\langle H_2^0 \rangle = 0,\hspace{0.5cm}
\langle H_3^0 \rangle  = 0,\hspace{0.5cm}
\langle H_4^0 \rangle  = 0.
\end{align}
In this way $\Phi_\alpha y_{ij}^{F\alpha}= y_{ij}^{F\alpha}R_{\alpha \beta}^T 
R_{\beta \gamma} \Phi_{\gamma}= \Y_{ij}^{F\beta} H_\beta $, and $F=U,D$;
where we have defined
\begin{align}
\label{eqC8}
\Y^{F\beta}_{ij}=R_{\beta \alpha}y_{ij}^{F\alpha}.
\end{align}
% By writing $\Y^{lF}_{ij}$ explicitly we can classify the different Two Higgs Doublet Model~(THDM) types
% \begin{align}\label{eq:mathcaly}
%  \Y^{1F}_{i,j}=& +\cos \beta y^{1F}_{i,j}+\sin \beta y^{2F}_{i,j}\notag\\
%  \Y^{2f}_{i,j}=& -\sin \beta y^{1F}_{i,j}+\cos \beta y^{2F}_{i,j}
% \end{align}
With these definitions equation~(\ref{eq:yukawa}) becomes
\begin{align}
\mathcal{L} = -\bar{q}_L^{\prime i} H_\beta \Y_{ij}^{D\beta}d^{\prime j}_{R}
              -\bar{q}_L^{\prime i} \tilde H_\beta \Y_{ij}^{U\beta}u^{\prime j}_{R}+\text{h.c}.
%              -\bar{l}_L^{\prime i} H_\beta \Y_{ij}^{E\beta}e^{\prime j}_{R}
%              -\bar{l}_L^{\prime i} \tilde H_\beta \Y_{ij}^{N\beta}\nu^{\prime j}_{R}   
\end{align}
It is necessary to rotate to the mass eigenstates of the fermion mass, i.e.,
\begin{align}
f_{L,R}= U^F_{L,R}f'_{L,R},
\end{align}
where the diagonalization matrices $U_{L,R}$  are defined below, in 
section~\ref{sec:mat-diag}.
From the Lagrangian for the charged currents 
\begin{align}
\mathcal{L}_{CC}=&-\frac{g}{\sqrt{2}} \bar{u}_{Li}'\gamma^{\mu} d_{Li}'W^+
%                  -\frac{g}{\sqrt{2}} \bar{e}_{Li}'\gamma^{\mu} \nu_{Li}'W^-
    +\text{h.c}\notag\\
                =&-\frac{g}{\sqrt{2}} \bar{u}_{Li}\gamma^{\mu}\left(V_{_{\text{CKM}}}\right)_{ij} d_{Lj}W^+
%                 -\frac{g}{\sqrt{2}} \bar{e}_{Li}\gamma^{\mu}V_{_{PMNS}} \nu_{Li}W^-
                 +\text{h.c},
\end{align}
it is possible to obtain the CKM mixing matrix  $V_{_{\text{CKM}}}= U^{U}_L U^{D \dagger}_L$ 
%and  $V_{_{PMNS}}= U^{E}_L U^{\nu \dagger}_L$ 
by rotating to the fermion mass eigenstates.
In particular, we are interested in the axial neutral current coupling to the axion
in the mass eigenstates 
\begin{align}\label{eq:yuk}
\mathcal{L}_{H^0} =&
              -\frac{1}{\sqrt{2}} \bar{d}_L^{\prime i}  H_\beta^0    \Y_{ij}^{D\beta}     d^{\prime j}_{R}
              -\frac{1}{\sqrt{2}}\bar{u}_L^{\prime i}  H_\beta^0 \Y_{ij}^{U\beta}     u^{\prime j}_{R}
%              -\bar{e}_L^{\prime i}   H_\beta^0    \Y_{ij}^{E\beta}     e^{\prime j}_{R}
%              -\bar{\nu}_L^{\prime i} H_\beta^{0*} \Y_{ij}^{N\beta}   \nu^{\prime j}_{R}
              +\text{h.c},\notag\\
              =&
              -\frac{1}{\sqrt{2}}\bar{d}_L^{i}  H_\beta^0    Y_{ij}^{D\beta}     d^{j}_{R}
              -\frac{1}{\sqrt{2}}\bar{u}_L^{i}   H_\beta^0    Y_{ij}^{U\beta}     u^{j}_{R}
%              -\bar{e}_L^{i}   H_\beta^0    Y_{ij}^{E\beta}     e^{j}_{R}
%              -\bar{\nu}_L^{i} H_\beta^{0*} Y_{ij}^{N\beta}   \nu^{j}_{R}
              +\text{h.c}, \notag\\
              \end{align}
where  $Y_{ij}^{F\beta} = \left(U^{F}_{L}   \Y^{F\beta}   U^{F\dagger}_{R}  
\right)_{ij} $.
In these expressions the mass functions in the interaction basis are:
\begin{align}\label{eq:b11}
M^{D}_{ij}= \frac{v}{\sqrt{2}} \Y_{ij}^{D1},\hspace{1cm}
M^{U}_{ij}= \frac{v}{\sqrt{2}} \Y_{ij}^{U1},
\end{align}
where $v= \langle H_1^0 \rangle$ is the Higgs vacuum expectation value. 

%%%%%%%%%%%%%%%%%%%%%%%%%%%%%%%%%%%%%%%%%%%%%%%%%%%%%%%%%%%%%%%%%%%%%%%%%%%%
%%%%%%%%%%%%%%%%%%%%%%%%%%%%%%%%%%%%%%%%%%%%%%%%%%%%%%%%%%%%%%%%%%%%%%%%%%%%
\section{Diagonalization matrices}
\label{sec:mat-diag}
In order to compare with physical quantities, it is necessary to rotate fields 
to the mass eigenstates, i.e.,
$u_{L,R}=U^{U}_{L,R}u'_{L,R}$ and $d_{L,R}=U^{D}_{L,R}d'_{L,R}$, where  prime 
means the interaction basis. 
In our formalism the mass matrices are Hermitian, hence 
the right-handed and left-handed diagonalizing matrices 
are identical; however, we obtain a minus sign 
on the quark mass eigenvalues of the second family (see comments after Eq.~\eqref{35b} and references~\cite{Branco:1999nb,Giraldo:2011ya}). To 
get a positive mass matrix  we introduce the 
identity matrix written as $I_2I_2=1$
with $I_2=\text{diag}(1,-1,1)$, i.e., 
\begin{align}\label{eq:c1}
M^{U}_{ij}=&\left(U^{U\dagger}\lambda^{U}U^{U}\right)_{ij}=
\left(U^{U\dagger}_L m^{U}U^{U}_R\right)_{ij}
=\frac{v}{\sqrt{2}}\mathcal{Y}^{U1}_{ij}\notag\\
=&\frac{v}{\sqrt{2}} 
R_{1\alpha}y^{U\alpha}_{ij},\notag\\
M^{D}_{ij}=&\left(U^{D\dagger}\lambda^{D}U^{D}\right)_{ij}
= \left(U^{D\dagger}_L m^{U}U^{D}_R\right)_{ij}
=\frac{v}{\sqrt{2}}\mathcal{Y}^{D1}_{ij}\notag\\
=& 
\frac{v}{\sqrt{2}}R_{1\alpha}y^{D\alpha}_{ij},
\end{align}
where:
\begin{align}
\lambda^{U,D}=&\text{diag}(m_{u,d},-m_{c,s},m_{t,b}),\notag\\
 m^{U,D}=&\text{diag}(m_{u,d},m_{c,s},m_{t,b}),\notag
 \end{align}
 \noindent the matrices $R$   
and   $\mathcal{Y}$ are defined in Eqs.~\eqref{eq:rab} and~\eqref{eqC8}, 
respectively; and 
\begin{align}\label{eq:url}
U_L^{U,D}= U^{U,D}, \hspace{1cm} U_R^{U,D}= I_2 U^{U,D},
\end{align}
where $U^{U,D}$ are the diagonalization matrices~\eqref{Uu} and \eqref{diagMd}.
In the second and fourth lines in~\eqref{eq:c1}  we made use of~\eqref{eq:b11}.
  It is important to stress that the texture-zeros pattern in the matrix 
$\mathcal{Y}^{F1}_{ij}$   are identical to those in the original Yukawa 
couplings $y^{F\alpha}_{ij}$, 
  since the sum over $\alpha$ does not mix the  $i, j$ indices.
  In fact, according to equations~\eqref{eq:yij}  and  \eqref{eq:yijd}, $M^{U,D}= 
\frac{v_{\alpha}}{\sqrt{2}}y^{U,D\alpha}_{ij}=\frac{v}{\sqrt{2}} 
R_{1\alpha}y^{U,D\alpha}_{ij}$, therefore
  $R_{1\alpha}=\frac{v_{\alpha}}{v}$. The diagonalization matrices are:
\begin{strip}  
{\footnotesize\begin{equation}
 U^{U\dagger}=
\begin{pmatrix}
e^{i (\phi_{C_u}+ \theta_{1u})} \sqrt{\frac{m_c m_t (A_u-m_u)}{A_u (m_c+m_u) 
(m_t-m_u)}} & -e^{i (\phi_{C_u}+ \theta_{2u})} \sqrt{\frac{(A_u+m_c) m_t 
m_u}{A_u (m_c+m_t) (m_c+m_u)}} & e^{i (\phi_{C_u}+ \theta_{3u})} \sqrt{\frac{m_c 
(m_t-A_u) m_u}{A_u (m_c+m_t) (m_t-m_u)}} \\
 -e^{i( \phi_{B_u}+ \theta_{1u})} \sqrt{\frac{(A_u+m_c) (m_t-A_u) m_u}{A_u 
(m_c+m_u) (m_t-m_u)}} & -e^{i (\phi_{B_u}+ \theta_{2u})} \sqrt{\frac{m_c 
(m_t-A_u) (A_u-m_u)}{A_u (m_c+m_t) (m_c+m_u)}} &e^{i (\phi_{B_u}+ \theta_{3u})} 
\sqrt{\frac{(A_u+m_c) m_t (A_u-m_u)}{A_u (m_c+m_t) (m_t-m_u)}} \\
 e^{i \theta_{1u}} \sqrt{\frac{m_u (A_u-m_u)}{(m_c+m_u) (m_t-m_u)}} & e^{i 
\theta_{2u}} \sqrt{\frac{m_c (A_u+m_c)}{(m_c+m_t) (m_c+m_u)}} & e^{i 
\theta_{3u}}\sqrt{\frac{m_t (m_t-A_u)}{(m_c+m_t) (m_t-m_u)}} 
\end{pmatrix},
\label{Uu}
\end{equation}}
{\footnotesize\begin{equation}
U^{D\dagger}=\begin{pmatrix}
 e^{i \theta_{1d}} \sqrt{\frac{m_b (m_b-m_s) m_s}{(m_b-m_d) (m_d+m_s) 
(m_b+m_d-m_s)}} & -e^{i \theta_{2d}} \sqrt{\frac{m_b (m_b+m_d) m_d}{(m_d+m_s) 
(m_b+m_d-m_s) (m_b+m_s)}} & \sqrt{\frac{m_d (m_s-m_d) m_s}{(m_b-m_d) 
(m_b+m_d-m_s) (m_b+m_s)}} \\
 e^{i \theta_{1d}} \sqrt{\frac{m_d (m_b-m_s)}{(m_b-m_d) (m_d+m_s)}} & e^{i 
\theta_{2d}} \sqrt{\frac{(m_b+m_d) m_s}{(m_d+m_s) (m_b+m_s)}} & \sqrt{\frac{m_b 
(m_s-m_d)}{(m_b-m_d) (m_b+m_s)}} \\
 -e^{i \theta_{1d}} \sqrt{\frac{m_d (m_b+m_d) (m_s-m_d)}{(m_b-m_d) (m_d+m_s) (m_b+m_d-m_s)}} & -e^{i \theta_{2d}} \sqrt{\frac{(m_b-m_s) m_s 
(m_s-m_d)}{(m_d+m_s) (m_b+m_d-m_s) (m_b+m_s)}} & \sqrt{\frac{m_b (m_b+m_d) 
(m_b-m_s)}{(m_b-m_d) (m_b+m_d-m_s) (m_b+m_s)}} 
\end{pmatrix},
\label{diagMd}
\end{equation}}
\end{strip}
where $\theta_{1u},\theta_{2u},\theta_{3u},\theta_{1d}$ y $\theta_{2d}$
are arbitrary phases~(a third phase for the diagonalization matrix 
~\eqref{diagMd} can be absorbed by the remaining phases) which are useful to 
adapt to the convention of the matrix 
$V_{\text{CKM}}=U^{U}_{L}\,U^{D\dag}_{L}$. 
Taking as input the SM parameters at the $Z$ pole, the best  fit values  are:

%\begin{table}[h]
\begin{strip}
\centering
\begin{tabular}{ccccccc}
\hline 
\hline 
$\theta_{1u}$&$\theta_{2u}$&$\theta_{3u}$&$\theta_{1d}$&$\theta_{2d}$& 
$\phi_{C_u}$&$ \phi_{B_u}$\\
\hline
$-2.84403$     &1.85606      &$-0.00461668$  &1.93013      &$-0.976639$    &
$-1.49697$    &0.301461\\
\hline
$A_u$      &$m_u$     &$m_c$      &$m_t$     &$m_d$      &$m_s$      &$m_b$ 
\\ 
\hline
1690.29~MeV&1.2684~MeV&633.197~MeV&171268~MeV&3.14751~MeV&56.1169~MeV&
2910.01~MeV\\  
%$T_L$       & 1/2& 3        &  0       & $x_{T_L}$ &    $x_{T_L}$   \\ 
\hline
\hline 
\end{tabular}
\captionof{table}{Best fit point of the mass matrices parameters to the quark masses and 
mixing angles at the $Z$ pole.}
\label{tab:eqF3}
\end{strip}
%\end{table}

%%%%%%%%%%%%%%%%%%%%%%%%%%%%%%%%%%%%%%%%%%%%%%%%%%%%%%%%%%%%%%%%%%%%%%%%%%%%
%%%%%%%%%%%%%%%%%%%%%%%%%%%%%%%%%%%%%%%%%%%%
\section{FCNC from~\texorpdfstring{$\Delta\mathcal{L}_{K^\psi}$}{}}
\label{sec:appendix-v-a-couplings}
The interaction term~\eqref{eq:deltaL} between the kinetic terms of the fermions and 
the axion is given by:
{\small \begin{align}
 \Delta\mathcal{L}_{K^\psi}=&
\frac{\partial_\mu a}{2f_a}\sum_{\psi}(x_{\psi_L}-x_{\psi_R})\bar{\psi}'\gamma^{\mu}\gamma^{5}\psi'\notag
                                     -(x_{\psi_L}+x_{\psi_R})\bar{\psi}\gamma^{\mu}\psi'\\
   =&
-\frac{\partial_\mu a}{2f_a}\sum_{\psi}x_{\psi_L}\bar{\psi}'\gamma^{\mu}(1-\gamma^{5})\psi'\notag
                                     +x_{\psi_R}\bar{\psi}\gamma^{\mu}(1+\gamma^5)\psi',
\end{align}}

By rotating from the interaction basis  to the mass basis  
for the SM quarks we obtain
\begin{align}
=&-\frac{\partial_\mu a}{2f_a}\Bigg(
    \bar{u}_{i}\gamma^{\mu}\left(1-\gamma^{5}\right)\Delta^{Uij}_{LL}(q)u_{j}\notag\\  
  +&  \bar{d}_{i}\gamma^{\mu}\left(1-\gamma^{5}\right)\Delta^{Dij}_{LL}(q)d_{j}\notag                         
 %%%%%%%%%%%%%%%%%%%%%%%%%%%%%%%%%%%%%%%%%%%%%%%%%%%%%%%%%%%%%%%%%%%%%%%%%%%%%%%%                                     
  + \bar{u}_{i}\gamma^{\mu}\left(1+\gamma^{5}\right)\Delta^{Uij}_{RR}(u)u_{j} \notag\\ +&  \bar{d}_{i}\gamma^{\mu}\left(1+\gamma^{5}\right)\Delta^{Dij}_{RR}(d)d_{j}\Bigg),%\notag\\
\end{align}
where {\small $\Delta^{Fij}_{LL}(q)= \left(U^D_{L}x_{q}~U_L^{D\dagger}\right)^{ij}$ } 
and   {\small $\Delta^{Fij}_{RR}(d)= \left(U^D_{R}x_{d}~U_R^{D\dagger}\right)^{ij}_.$}
From these expressions we are interested in the terms:
 {\footnotesize
\begin{align}\label{eq:LKD}
 \Delta\mathcal{L}_{K^D}=&
-\frac{\partial_\mu a}{2f_a}\Bigg(
     \bar{d}_{i}\gamma^{\mu}\left(1-\gamma^{5}\right)\Delta^{Dij}_{LL}(q)d_{j}\notag\\      
  +&  \bar{d}_{i}\gamma^{\mu}\left(1+\gamma^{5}\right)\Delta^{Dij}_{RR}(d)d_{j}\Bigg)\\
  =& 
-\frac{\partial_\mu a}{2f_a}
     \bar{d}_{i}\gamma^{\mu}\left(\Delta^{Dij}_{V}                         
  +  \gamma^{5}\Delta^{Dij}_{A}\right)d_{j},\notag
  \notag\\
  =& -\partial_\mu a
     \bar{d}_{i}\gamma^{\mu}\left(g_{af_if_j}^{V}                         
  +  \gamma^{5}g_{af_if_j}^{A}\right)d_{j},
  \end{align}}
  
%\subsection{ V-A effective couplings  and the Meson decay width}
%\label{sec:appendix-v-a-couplings}

 %
From this expression, we can infer vector and axial couplings for any type of fermions $F=U,D,E,N$
 \begin{align}\label{eq:gva}
 g_{af_if_j}^{V,A}=
 \frac{1}{2f_a}\Delta^{Fij}_{V,A}.
 \end{align}
 
These couplings~\eqref{eq:gva}, generate FCNC processes as those shown in 
Fig.~\ref{fig:ktopi}.
% \appendix
% \section{\texorpdfstring{$\Gamma\left(K^+\longrightarrow \pi^+ a\right)$}{}}
 According to reference~\cite{Griffiths:2008zz}
 \begin{align}\label{eq:decay}
 \Gamma = \frac{S\lvert \vec{p} \rvert}{8\pi m_K^2}
 \lvert \mathcal{M} \rvert^2, 
 \end{align}
 where $\lvert \vec{p} \rvert= m_K\lambda_{K\pi a}^{1/2}/2$, and
 
\noindent $\lambda_{K\pi a}= \left(1-\frac{(m_a+m_\pi)^2}{m_K^2}\right)\left(1-\frac{(m_a-m_\pi)^2}{m_K^2}\right)$  and $S=1$. 
 The leading order $S$ matrix element    for $K^-\rightarrow \pi^-a$ is
 {\footnotesize
 \begin{align}
   \mathcal{M}=& \langle \pi^-(p_{\pi}),a(p_a)\lvert i\mathcal{L}(s\rightarrow d)\lvert K^-(p_{K})\rangle\notag\\
  =&  - ig_{ads}^V(p_K-p_\pi)_{\mu}
  \langle \pi^-(p_{\pi})\lvert \bar{d}\gamma^\mu s\lvert K^-(p_{K})\rangle
  \langle a(p_a)\lvert a(p_a)\rvert 0\rangle   
  \notag\\
  =& - ig_{ads}^V(m_K^2-m_\pi^2)f_0(q^2), 
 \end{align}}
 where $q^{2}=(p_K-p_\pi)^2$ and : 
 \begin{eqnarray}
  f_{0}(q^2)&=&f_{+}(q^2) +\frac{q^2f_{-}(q^2)}{(m_K^2-m_\pi^2)} ,\\
  \langle a(p_a)\lvert a(p_a)\rvert 0\rangle&=&\langle a(p_a)|a(p_a)\rangle=1
 \end{eqnarray}

 As the initial and final states have the same parity only 
 the matrix elements of the vector current are different from zero~\cite{Langacker:2010zza}, 
 then 
  \begin{align}
 \Gamma(K^+\rightarrow \pi^+a) = \frac{m_K^3}{16\pi}\left(1-\frac{m_\pi^2}{m_K^2}\right)^2\lambda_{K\pi a}^{1/2} f^2_0(m_a^2)
 \lvert g_{ads}^V \rvert^2.
  \end{align}
To calculate the $B\rightarrow V a$ decay width, 
where $V$ is a vector meson,
it is necessary to consider the form factors for the 
quark level process $b\rightarrow q$~\cite{Horgan:2013hoa}

{\small
\begin{align}
\langle V(k,\epsilon)\lvert \bar{q}\gamma^{\mu}b\rvert B(p)\rangle
=& \frac{2iV(q^2)}{m_B+m_V}\epsilon^{\mu\nu\rho\sigma}\epsilon^{*}_{\nu}k_{\rho}p_{\sigma},\\
\langle V(k,\epsilon)\lvert \bar{q}\gamma^{\mu}\gamma^{5}b\rvert B(p)\rangle
=&  2m_V A_{0}(q^2)\frac{\epsilon^*\cdot q}{q^2}q^{\mu}\notag\\
+&(m_B+m_V)A_{1}(q^2)\left(\epsilon^{*\mu}
-\frac{\epsilon^*\cdot q}{q^2}q^{\mu}\right)
\notag\\
-&\frac{A_{2}(q^2)\epsilon^*\cdot q}{(m_b+m_V)}
\left[(p+k)^{\mu}-\frac{m_B^2-m_V^2}{q^2}q^{\mu}\right].
\end{align}}

There are also strong constraints from 
the decay $B\rightarrow K^*a$, the $K^*$ kaon is a vectorial meson, parity-even under inversion of the spatial coordinates. 
Due to the selection rules of the Lorentz group only the 
axial-vector matrix elements $\langle K^* \lvert \bar{s}\gamma^{\mu}\gamma^5  b\rvert B \rangle$
are different from zero 
 \begin{align}
  \mathcal{M}=& -ig_{asb}^Vq_{\mu}
  \langle K(p_{K})\lvert \bar{s}\gamma^\mu\gamma^5 b\lvert B(p_{P})\rangle\notag\\
  =&  -ig_{asb}^V
 2m_{K^*} A_{0}(q^2)\epsilon^*\cdot q, 
  \notag
 \end{align}
where $q_{\mu}=(p_B-p_K)_{\mu}$.
Summing over the final polarization states
 $\sum_{s}\epsilon^{\mu *}(s)\epsilon^{\nu *}(s)=\left(-g^{\mu\nu}+\frac{p_{K^*}^{\mu}p_{K^*}^{\nu}}{m_{K^*}^2}\right)$, we get 
\begin{align}
 \sum_{s}\lvert \mathcal{M} \rvert^2
  =
   \lvert g_{asb}^A\rvert^2 A_{0}^2(m^2_a) 
  m_B^4\lambda_{BK^{*}a},
\end{align}
and replacing this result in Eq.~\ref{eq:decay} the width decay can be written as
 \begin{align}
 \Gamma(B\rightarrow K^{*}a) = \frac{m_B^3}{16\pi}
 \lambda_{BK^{*}a}^{3/2} A_{0}^2(m^2_a)\lvert g_{asb}^A\rvert^2. 
 \end{align}

%%%%%%%%%%%%%%%%%%%%%%%%%%%%%%%%%%%%%%%%%%%%%%%%%%%%%%%%%%%%%%%%%%%%%%%%%%%%
\section{Scalar potential}
\label{sec:scalars}
In order to explain the textures of the mass matrices of our model, four scalar doublets $\Phi_\alpha$ were introduced in section~\ref{sec:III}, additionally a scalar singlet $S$ is required to break the  PQ symmetry. For completeness it is necessary to introduce a potential $V(\Phi,S_1,S_2)$ with all the terms  allowed by the PQ symmetry. From this potential it is possible to obtain the masses of the scalar fields allowing us to determine which of them correspond to  Goldstone bosons.  One of the CP odd massless  scalars must correspond to the  axion field associated with the breaking of PQ symmetry.   
The most general CP invariant scalar potential in the PQ symmetry scenario is
%
%\begin{widetext}
{\footnotesize
\begin{eqnarray}\label{eq:scalar-potential}
V(\Phi,S_i) &=& \sum_{i=1}^4\mu_{i}^{2}\Phi_{i}^{\dagger}\Phi_{i} +\sum_{k=1}^2\mu_{s_k}^2 S_k^{*}S_k+\sum_{i=1}^4\lambda_{i}\left(\Phi_{i}^{\dagger}\Phi_{i}\right)^{2}\nonumber \\
&+&  \sum_{k=1}^2\lambda_{s_k}\left(S_k^{*} S_k \right)^{2}     + \sum_{i=1}^4\sum_{k=1}^2\lambda_{is_k}\left(\Phi_{i}^{\dagger}\Phi_{i}\right)\left(S_k^{*}S_k\right)\nonumber\\
&+& \sum_{\underbrace{i,j=1}_{i<j}}^4\bigg{(}\lambda_{ij}\left(\Phi_{i}^{\dagger}\Phi_{i}\right) \left(\Phi_{j}^{\dagger}\Phi_{j}\right)+J_{ij}\left(\Phi_{i}^{\dagger}\Phi_{j}\right) \left(\Phi_{j}^{\dagger}\Phi_{i}\right)\bigg{)}
\nonumber\\
&+& \lambda_{s_1 s_2}\left(S_1^{*}S_1\right)\left(S_2^{*}S_2\right)\nonumber\\
&+&K_{1}\left(\left(\Phi_{1}^{\dagger}\Phi_{2}\right) \left(\Phi_{3}^{\dagger}\Phi_{2}\right) + h.c.\right)\nonumber\\
&+&
 K_{2}\left(\left(\Phi_{3}^{\dagger}\Phi_{4}\right) \left(\Phi_{3}^{\dagger}\Phi_{1}\right) + h.c.\right)\nonumber\\&+& F_1 \left( \left(\Phi_{2}^{\dagger}\Phi_{3}\right) S_1 +h.c.\right)\nonumber\\
&+&F_2 \left( \left(\Phi_{1}^{\dagger}\Phi_{2}\right) S_1 +h.c.\right)\nonumber\\
&+&\frac{1}{2} \left(m_{\zeta_{S_2}}\right)^2_{\text{SB}}\zeta^2_{S_2}+\frac{1}{2}\left(m_{\xi_{S_2}}\right)^2_{\text{SB}}\xi^2_{S_2}.
\end{eqnarray}}
%\end{widetext}

%
where the terms proportional to $F_i$ are allowed by the particular choice of the PQ charges and  the  $F_i$ couplings  have units of mass.  
After spontaneous symmetry breaking~(SSB), the four Higgs doublets acquire a VEV that gives masses to all SM particles and the scalar doublets could be written as
\begin{align}
\label{eq:higgs1}
\Phi_\alpha =& 
\begin{pmatrix}
\phi_\alpha^{+}\\
\frac{v_\alpha+h_\alpha+i\eta_\alpha}{\sqrt{2}}
\end{pmatrix},
\hspace{1cm}
\tilde\Phi_\alpha=i\sigma_2 \Phi_\alpha^{*},
\hspace{0.5cm}\notag\\
S_i=&\frac{v_{_{S_i}}+\xi_{S_i}+i\zeta_{S_i}}{\sqrt{2}};\hspace{1cm} i=1,2.
\end{align}
The singlet scalar field $S_1$ breaks the PQ symmetry at the high energy scale given by $v_{s_1}$. 
The last two terms in the equation \eqref{eq:scalar-potential} correspond to soft breaking masses of the  imaginary and the real part of $S_2$,
which are generated at one  loop in the Coleman-Weinberg potential from the interaction term $\lambda_Q S_2\bar{Q}_RQ_L+\text{h.c.}$
From Eqs.~\eqref{eq:vevs} and~\eqref{eq:vevs2} we have the following hierarchy among VEVs, $v_4\ll v_1, v_2\ll v_3\ll v_{S_1}\sim v_{S_2}$.
In the scalar sector, we  have CP-even, CP-odd, and charged fields.
As shown in Appendices~\ref{sec:cpeven}, \ref{sec:chargued} and  \ref{sec:cpodd} by choosing the couplings close to one, as follows:
\begin{eqnarray}\label{eq:couplings}
\lambda_{1}&=& \lambda_{2}=\lambda_{4}=\lambda_{s_1}=\lambda_{s_2}=\lambda_{s_1 s_2}=
1,\hspace{1cm}\nonumber\\
\lambda_3&=&0.463\nonumber\\
\lambda_{ij}&=&1 \textup{ for any } i,j,  \nonumber\\
\lambda_{js_1}&=&\lambda_{js_2}=1\textup{ for any }j,\nonumber\\
J_{12}&=& J_{13}=J_{23}=J_{24}=-1,\ \  \textup{otherwise } J_{ij}=1,\nonumber\\
K_1&=&K_2=-1,\nonumber\\
F_1&=&F_2=-1 \text{GeV},
\end{eqnarray}

we obtain scalar masses above the TeV scale (except for the Higgs boson) allowing them to avoid  LHC constraints on heavy Higgs bosons~\cite{ATLAS:2019tpq} and charged scalar bosons~\cite{CMS:2019rlz}.
The $\lambda_{3}$ value was chosen in order to adjust the SM Higgs mass.
%As shown in Appendix~\ref{sec:cpeven}  the square of the SM Higgs mass is given by
%\begin{align}
%m_h^2 = v_3^2\left(2\lambda_{3}-\frac{2\lambda_{3s_1}
%}{3\lambda_{s_1}}\right)
%\end{align}
In our approach, the $v_i$~(Eqs.~\eqref{eq:vevs}  and~\eqref{eq:vevs2}) are determined from the SM fermion masses and the quark mass matrix texture,  $v_{s_1}$ is still a free parameter, nonetheless, this parameter is important for the axion physics due to  the relation~\cite{Giannotti:2017hny},
\begin{align}
f_a=\frac{v_{s_1}}{2N}.
\end{align}
In our calculations we took $v_{s_1}\approx v_{s_2}\approx 10^{6}$GeV. It is important to emphasize that in our model $ f_a $ can take arbitrary values, however, a small $f_a$ restrict $\epsilon$ (Eq.~\ref{eq:parametrization}) to values close to zero.
Taking into account all these considerations and the Eq.~\eqref{eq:couplings} the scalar mass spetrum~(in GeV) is:
{\footnotesize
\begin{align}
&\text{CP even}         =\{1.73\times10^6,1.\times10^6,6.54\times10^3,1.97\times10^3,\nonumber\\
&\hspace{1.8cm}1.09\times 10^3,125\},\nonumber\\
&\text{CP odd}          =\{6.54\times10^3,1.97\times10^3,1.09\times 10^3,0,0,m_{\zeta_{S_2}}\},\nonumber\\
&\text{Charged fields} = \{6.54\times10^3,1.97\times10^3,1.11\times 10^3,0\}.
\end{align}}
The mass spectrum of the scalar fields is above the TeVs scale, except the SM Higgs  which was set to 125~GeV.
The pseudoscalar sector (CP odd fields) have two zero mass eigenstates, the axion field  and the Goldstone boson 
which is absorbed by the longitudinal component of the SM $ Z $ boson.  A similar result is achieved in the charged sector where it is possible to identify the two Goldstone bosons needed to give mass to the SM  $W^{\pm}$ fields.
%Approximate  analytical expressions for the mass matrices and their eigenvalues are shown in Appendix~\ref{sec:scalars}.
%%%%%%%%%%%%%%%%%%%%%%%%%%%%%%%%%%%%%%%%
%\section{Scalar mass matrices}
%\label{sec:scalars}
\subsection{CP-even scalar sector\label{sec:cpeven}}
As shown in Eq.~\ref{eq:higgs} after the SSB the four Higgs doublets and the scalar singlets 
acquire VEVs, yielding the squared mass matrix $M_{R}^{2}$ for CP-even scalar particles expressed in the $(h_{1},h_{2},h_{3},h_{4},\xi_{S_1}, \xi_{S_2})$ basis, with entries given by:
\begin{eqnarray}
M_{R_{11}}^2&=&-\frac{K_1v_2^2 v_3+K_2 v_3^2 v_4+\sqrt{2} F_2 v_2 v_{s_1}-4 v_1^3\lambda_1}{2 v_1},\nonumber\\
M_{R_{12}}^2&=&\frac{F_2 v_{s_1}}{\sqrt{2}}+v_2 (K_1v_3+v_1 H_{12}),\nonumber\\
M_{R_{13}}^2&=&\frac{K_1v_2^2}{2}+v_3 (K_2 v_4+v_1 H_{13}),\nonumber\\
M_{R_{14}}^2&=& \frac{K_2 v_3^2}{2}+v_1 v_4 H_{14},\nonumber \\
M_{R_{15}}^2&=&\frac{F_2 v_2}{\sqrt{2}}+v_1 v_{s_1} \lambda_{1s_1},\nonumber \\
M_{R_{22}}^2&=&  \frac{-\sqrt{2} (F_2 v_1+F_1 v_3) v_{s_1}+4 v_2^3 \lambda_2}{2 v_2},\nonumber\\
M_{R_{23}}^2&=&K_1v_1 v_2+\frac{F_1 v_{s_1}}{\sqrt{2}}+v_2 v_3 H_{23} \nonumber\\
M_{R_{24}}^2&=&v_2 v_4 H_{24}, \nonumber\\
M_{R_{25}}^2&=& \frac{F_2 v_1}{\sqrt{2}}+\frac{F_1 v_3}{\sqrt{2}}+v_2 v_{s_1} \lambda_{2s_1},\nonumber\\
M_{R_{33}}^2&=&-\frac{K_1v_1 v_2^2+\sqrt{2} F_1 v_2 v_{s_1}-4 v_3^3 \lambda_{3}}{2 v_3},\nonumber\\
M_{R_{34}}^2&=& v_3 (K_2 v_1+v_4 H_{34}),\nonumber\\
M_{R_{35}}^2&=&\frac{F_1 v_2}{\sqrt{2}}+v_3 v_{s_1} \lambda_{3S_1},\nonumber \\
M_{R_{44}}^2&=&-\frac{K_2 v_1 v_3^2}{2 v_4}+2 v_4^2 \lambda_{4},\nonumber \\
M_{R_{45}}^2&=& v_4 v_{s_1} \lambda_{4s_1},\nonumber\\
M_{R_{55}}^2&=&\frac{-\sqrt{2} v_2 (F_2 v_1+F_1 v_3)+4 v_{s_1}^3\lambda_{S_1}}{2 v_{s_1}}\nonumber\\
M_{R_{j6}}^2&=&\lambda_{js_2}v_jv_{s_2}\text{, for  j<5 } \nonumber \\
M_{R_{56}}^2&=&\lambda_{s_1s_2}v_{s_1}v_{s_2}\nonumber \\
M_{R_{66}}^2&=&2\lambda_{s_2}v_{s_2}^2+m_{\xi_{S_2}}^2.
\end{eqnarray}

where $H_{ij}=\lambda_{ij}+J_{ij}$.  At  leading order, the eigenvalues of this matrix are approximately (in GeV):
{\small
\begin{align}
%&\bigg\{-\frac{K_2 v_1 v_3^2}{2 v_4},-\frac{F_2 v_2 v_{s_1}}{\sqrt{2} v_1},-\frac{F_1 v_3 v_{s_1}}{\sqrt{2} v_2},2\lambda_{3}v_3^2-\frac{2\lambda_{3s_1}v_3^2
%}{3\lambda_{s_1}}
%,\nonumber\\
%&\hspace{1cm}3 \lambda_{s_1} v_{s_1}^2, \lambda_{s_1} v_{s_1}^2\bigg\}.
&\{1.73\times10^6,1.\times10^6,6.54\times10^3,1.97\times10^3,\nonumber\\
& 1.09\times 10^3,125\},
\end{align}}

Where the  value  $125$~GeV   corresponds to the SM Higgs.
Hereafter, the  signs of the couplings are chosen in such a way that the eigenvalues are positive.
\subsection{Charged scalar sector\label{sec:chargued}}

The square of the mass matrix for the charged scalar sector, $M_C^2$,  after SSB in the  Higgs sector, can be written in the
$(\phi_{1}^\pm, \phi_{2}^\pm, \phi_{3}^\pm, \phi_{4}^\pm)$ basis as

\begin{eqnarray}
M_{C_{11}}^2&=&-\frac{J_{12} v_1 v_2^2+K_1v_2^2 v_3+J_{13} v_1 v_3^2}{2 v_1}\nonumber\\
&+&\frac{K_2 v_3^2 v_4+J_{14} v_1 v_4^2+\sqrt{2} F_2 v_2 v_{s_1}}{2 v_1},\nonumber\\
M_{C_{12}}^2&=& \frac{1}{2} \left(J_{12} v_1 v_2+K_1v_2 v_3+\sqrt{2} F_2 v_{s_1}\right),\nonumber\\
M_{C_{13}}^2&=&\frac{1}{2} v_3 (J_{13} v_1+K_2 v_4),\nonumber  \\
M_{C_{14}}^2&=& \frac{J_{14} v_1 v_4}{2},\nonumber\\
M_{C_{22}}^2&=&-\frac{J_{12} v_1^2 v_2+2 K_1v_1 v_2 v_3}{2 v_2}\nonumber\\
&+&\frac{J_{23} v_2 v_3^2+J_{24} v_2 v_4^2+\sqrt{2} F_2 v_1 v_{s_1}+\sqrt{2} F_1 v_3 v_{s_1}}{2 v_2}, \nonumber\\
M_{C_{23}}^2&=&\frac{1}{2} \left(K_1v_1 v_2+J_{23} v_2 v_3+\sqrt{2} F_1 v_{s_1}\right),\nonumber \\
M_{C_{24}}^2&=&\frac{J_{24} v_2 v_4}{2},\nonumber \\
M_{C_{33}}^2&=& -\frac{K_1v_1 v_2^2+J_{13} v_1^2 v_3+J_{23} v_2^2 v_3}{2 v_3}\nonumber\\
&+&\frac{2 K_2 v_1 v_3 v_4+J_{34} v_3 v_4^2+\sqrt{2} F_1 v_2 v_{s_1}}{2 v_3},\nonumber \\
M_{C_{34}}^2&=& \frac{1}{2} v_3 (K_2 v_1+J_{34} v_4),\nonumber \\
M_{C_{44}}^2&=& -\frac{K_2 v_1 v_3^2+J_{14} v_1^2 v_4+J_{24} v_2^2 v_4+J_{34} v_3^2 v_4}{2 v_4}.
\end{eqnarray}
For this matrix,  eigenvalues numerically are  close  its diagonal elements, i.e., (in GeV): 
\begin{align}
%\approx
%\left\{
%-\frac{\sqrt{2} F_2 v_2 v_{s_1}}{2 v_1}, 
%-\frac{\sqrt{2} F_1 v_3\,v_{s_1}}{2 v_2},
%-\frac{K_2\,v_1\,v_3^2}{2 v_4},
%0\right\}
\{6.54\times10^3,1.97\times10^3,1.11\times 10^3,0\}.
\end{align}
The zero mass eigenvalue corresponds to the Goldstone Boson absorbed in the longitudinal component of the charged vector fields $W^\pm$.
%
%%%%%%%%%%%%%%%%%%%%%%
%%%%%%%%%%%%%%%%%%%%%%
\subsection{CP-odd scalar sector\label{sec:cpodd}}
The  square of the mass matrix for the CP-odd scalars in the $(\eta_{1}, \eta_{2}, \eta_{3}, \eta_{4}, \zeta_{S_1}, \zeta_{S_2})$ basis is given by
\begin{eqnarray}
M_{I_{11}}^2&=&  -\frac{K_1v_2^2 v_3+K_2 v_3^2 v_4+\sqrt{2} F_2 v_2 v_{s_1}}{2 v_1}, \nonumber  \\
M_{I_{12}}^2&=&  K_1v_2 v_3+\frac{F_2 v_{s_1}}{\sqrt{2}},\nonumber  \\
M_{I_{13}}^2&=&  -\frac{K_1v_2^2}{2}+K_2 v_3 v_4,\nonumber \\
M_{I_{14}}^2&=& -\frac{K_2 v_3^2}{2}, \nonumber \\
M_{I_{15}}^2&=&  \frac{F_2 v_2}{\sqrt{2}},\nonumber  \\
M_{I_{22}}^2&=& -\frac{4 K_1v_1 v_2 v_3+\sqrt{2} (F_2 v_1+F_1 v_3) v_{s_1}}{2 v_2},\nonumber\\
M_{I_{23}}^2&=&  K_1v_1 v_2+\frac{F_1 v_{s_1}}{\sqrt{2}},\nonumber \\
M_{I_{24}}^2&=& 0,\nonumber  \\
M_{I_{25}}^2&=& \frac{-F_2 v_1+F_1 v_3}{\sqrt{2}},\nonumber   \\
M_{I_{33}}^2&=& -2 K_2 v_1 v_4-\frac{v_2 \left(K_1v_1 v_2+\sqrt{2} F_1 v_{s_1}\right)}{2 v_3},\nonumber\\
M_{I_{34}}^2&=&  K_2 v_1 v_3,\nonumber \\
M_{I_{35}}^2&=&  -\frac{F_1 v_2}{\sqrt{2}},\nonumber   \\
M_{I_{44}}^2&=& -\frac{K_2 v_1 v_3^2}{2 v_4},\nonumber  \\
M_{I_{45}}^2&=& 0,\nonumber \\
M_{I_{55}}^2&=& -\frac{v_2 (F_2 v_1+F_1 v_3)}{\sqrt{2} v_{s_1}}\nonumber\\
M_{I_{j6}}^2&=&0\text{, for  j<6} \nonumber\\
M_{I_{66}}^2&=&m_{\zeta_{S_2}}^2
\end{eqnarray}
At leading order the eigenvalues of this matrix are (in GeV):
\begin{align}
%\approx \left\{0,0,0,-\frac{K_2 v_1 v_3^2}{2 v_4},-\frac{F_2 v_2 v_{s_1}}{\sqrt{2} v_1},-\frac{F_1 v_3 v_{s_1}}{\sqrt{2} v_2}\right\}.
\{6.54\times10^3,1.97\times10^3,1.09\times 10^3,0,0,m_{\zeta_2}\}
\end{align}
One of the  zero mass eigenvalues corresponds to the Goldstone Boson absorbed in the longitudinal component of the neutral vector field of the SM  $Z^{0}$, and the other one corresponds to the axion field $a$.

%%%%%%%%%%%%%%%%%%%%%%%%%%%%%%%%%%%%%%%%%%%%%%%%%%%%%%%%%%%%%%%%%%%%%%%%%%%%
%
\bibliographystyle{ieeetr}
\bibliography{bibliografia}

\end{document}